\documentclass[11pt]{article}
\pdfoutput=1
\usepackage{jcappub}
\usepackage{array}
\usepackage{github_aas_macros}
\usepackage{graphicx}
\usepackage{epsfig}
\usepackage{float}
\usepackage{subcaption}

\usepackage[normalem]{ulem}
\newcommand{\be}{\begin{equation}}
\newcommand{\ee}{\end{equation}}

\newcommand{\mpch}{\, {\rm Mpc/h}}

\newcommand{\ihmpc}{\, {\rm h/Mpc}}

\def\bi#1{\hbox{\boldmath{$#1$}}}

\newcommand{\fig}[1]{Figure \ref{#1}}

\title{Cosmological Reconstruction From Galaxy Light: Neural Network Based Light-Matter Connection}  

\author[a]{Chirag Modi,}
\author[a]{Yu Feng,}
\author[a,b]{Uro\v s Seljak}

\affiliation[a]{Berkeley Center for Cosmological Physics and Department of Physics, University of California, Berkeley, CA 94720}
\affiliation[b]{Physics Division, Lawrence Berkeley National Laboratory, Cyclotron Rd, Berkeley, CA 94720}

\emailAdd{modichirag@berkeley.edu, yfeng1@berkeley.edu, useljak@berkeley.edu}

\abstract{We present a method to reconstruct the initial conditions of the universe using observed galaxy positions and luminosities under the assumption that the luminosities can be calibrated with weak lensing to give the mean halo mass. Our method relies on following the gradients of forward model and since the standard way to identify halos is non-differentiable and results in a discrete sample of objects, we propose a framework to model the halo position and mass field starting from the  non-linear matter field using Neural Networks (NN), which are differentiable, yet can produce very pointlike maps.
We evaluate the performance of our model with multiple metrics and find that our model is more than $95\%$ correlated with the halo-mass fields up to $k\sim 0.7 \ihmpc$, and significantly reduces the stochasticity over the Poisson shot noise.
We develop a data likelihood model that takes our modeling error and intrinsic scatter in the halo mass-light relation into account and show that a 
displaced log-normal model is a good approximation to it.
We optimize over the corresponding loss function to reconstruct the initial density field of the dark matter starting from the halo mass field. To speed up and improve the convergence, we develop an annealing procedure for several parameters in the loss function, such as smoothing the likelihood starting with large smoothing and gradually decreasing it. We apply the method to  
halo number densities of $\bar{n} = 2.5\times 10^{-4}\ -\ 10^{-3}({\rm h/Mpc})^3$, typical of current and future redshift surveys, and recover a Gaussian initial density field, mapping all the higher order information in the data into the power spectrum of the linear reconstructed field.
We show that our reconstruction improves over the standard reconstruction.  For baryonic acoustic oscillations (BAO) the gains are relatively modest because BAO is dominated by large scales where standard reconstruction suffices. We improve upon it by $\sim 15-20\%$ in terms of error on BAO peak as estimated by Fisher analysis at $z=0$.
We expect larger gains will be achieved when applying this method to the broadband linear power spectrum reconstruction on smaller scales.  
}

\begin{document}
\maketitle
\flushbottom

\section{Introduction}
\label{sec:intro}
Studies of the large scale structure (LSS) of the Universe have played a very important role in establishing the standard model
of cosmology.
As a community, we are also investing a large amount of resources into the current and upcoming large scale imaging surveys such as the 
Dark Energy Survey (DES\footnote{\url{https://www.darkenergysurvey.org/}}), 
Dark Energy Spectroscopic Survey (DESI\footnote{\url{http://desi.lbl.gov/}}), 
Large Synoptic Survey Telescope (LSST\footnote{\url{https://www.lsst.org}}), Euclid\footnote{\url{http://sci.esa.int/euclid}} and 
WFIRST\footnote{\url{https://wfirst.gsfc.nasa.gov}}.
Due to the sheer number of modes that can be observed in the three-dimensional map of the Universe, there is a vast amount of statistical information about the initial conditions of the Universe and the cosmological parameters available to be extracted from these surveys. Furthermore, most of this information is present on smaller scales, since the number of modes scales as $k_{max}^3$, where $k_{max}$ is the smallest wavevector. 

However, the statistics of small scales in the late time Universe are complicated to model due to the highly nonlinear gravitational evolution. A large portion of the cosmological information is at best 
encoded into the higher order statistics, or at worst lost due to noise, and hence these scales are generally excluded from the cosmological analysis.
In an effort to recover more of this information, the issue of `reconstruction' 
of initial conditions has received a lot of attention over the last decade. The hope is to partly undo non-linear evolution of the Universe and recover the initial conditions, which are thought to be Gaussian and hence can then be described using a power spectrum statistics alone which contains
all the information (apart from parameters which control the gravitational evolution).

As a result, different flavors of reconstruction have been proposed over the years. One of the most successful methods, now referred to as standard reconstruction, was proposed in \cite{Eisenstein2007} wherein one estimates the linear density field by reversing the Zeldovich displacement in the clustered (galaxy) and random catalog, as estimated from filtered non-linear density. Recently, similar methods with improved estimates of non-linear displacement have been developed \cite{Zhu2017,Schmittfull2017}. 
A conceptually different approach to reconstruction involves sampling initial density modes with Hamiltonian Monte Carlo (HMC), which are then evolved using a forward model of choice (N-body, PM, 2LPT) and compared against the true non-linear density field \cite{Jasche2013,Kitaura2013, Wang2014}.
An alternate approach is to instead reconstruct the initial density field by optimizing the posterior of initial density modes, followed by analytic marginalization over these modes to reconstruct the summary statistics that optimally contain the desired information \cite{Seljak2017}. A common feature of the latter two (sampling and optimization) approaches is that they require knowledge of the gradient of the forward data model with respect to the initial modes to perform efficiently.

In this work, we are motivated to apply these reconstruction techniques to large scale imaging surveys and make use of the wealth of data provided by them. These surveys map the skies in several wavelength bands to get the positions of the galaxies as well as an estimate of their stellar mass and bolometric luminosity. This means that they do not map the dark matter field directly but instead observe a biased tracer (galaxies), along with the derived properties  (stellar mass or luminosity) associated with the dark matter mass that may be prone to statistical and systematic noise.
To be able to reconstruct initial density field from these observables, one needs to be able to forward model to these quantities directly.

In traditional cosmological analysis, this is achieved in N-Body and Particle Mesh (PM) simulations by first identifying dark matter groups (halos) using algorithms such as Friends of friends (FOF) or Spherical Overdensity halo finders.
These are then populated with galaxies using techniques such as Halo Occupation Distribution (HOD) \cite{Berlind2002} and assigned luminosity and stellar mass based on some relationship with the parent halo's mass. 
For the purpose of reconstruction, this approach has proven to be a complication, especially for the methods that rely on being able to take the gradient of the forward model: the derivatives of FOF and other halo finding or galaxy painting algorithms such as HOD with respect to the initial modes can not be easily evaluated. Moreover, the halo field is discrete and for discrete fields, it is difficult to write a suitable loss function. 

Several ways have been proposed to overcome this problem.
The simplest way to relate matter field to galaxy field is to use a bias model, as is done in standard reconstruction \cite{Eisenstein2007}. Linear bias model is also used in \cite{Jasche2013} to relate matter to the halo field, but they treat galaxies as a Poisson sampling of this biased field and develop the likelihood function to take this sampling into account. On an object by object basis this approach is unlikely to be very accurate since it does not reduce stochasticity over the Poisson shot noise. 
Another approach involves backward modeling from the galaxy catalogs to `reconstruct' a nonlinear matter field first, and then using this field to do a reconstruction of initial conditions. For example, \cite{Wang2009} has proposed a method to assign matter profiles to halos which have been constructed from simulations and \cite{Wang2013} has used this matter field to perform initial condition reconstruction.
Alternatively, \cite{Yu2017} performs a Delaunay tessellation \cite{Cautun2011} of dark matter halos to reconstruct the continuous nonlinear matter field.
Since these are backward modeling approaches it is difficult to insert modeling uncertainties into the process, and it is also difficult to develop a reliable noise model: typically noise is only uncorrelated in the data space (where it is simply detector noise or equivalent), and any backward modeling correlates the latent space of variables (such as the nonlinear dark matter density) in a way that would need to be modeled, but becomes very expensive to do so. Moreover, nonlinear dark matter is not our ultimate goal: 
what we want is to map higher order information (e.g. bispectrum and higher order statistics, various 
peak, filament and void statistics etc.) into two point function statistic, since this leads to an 
optimal extraction of information from the data \cite{Seljak2017}. 

In this work, we propose an alternate approach: we develop a forward model to go from the linear to the nonlinear matter field and then to the halos, galaxies and their corresponding observable properties in a \textit{differentiable} fashion, such that its gradient can be evaluated and used for reconstruction. To make it differentiable we have to give up the concept of a pointlike galaxy or halo.
However, forward models are preferable to backward models since the loss function can be written in the data space, where the noise is diagonal. This also makes it easier to incorporate some of the most prominent uncertainties and systematics in LSS analysis such as survey geometry and selection effects, incompleteness due to fiber collisions etc.
Finally, the forward models can easily incorporate modeling uncertainties, such as relation between galaxy luminosity and halo mass, various satellite statistics etc. In the forward model, one parametrizes the effect of these nuisance variables on the data and then performs statistical analysis of these parameters together with the cosmological parameters. 

\begin{figure}
\centering
\noindent\makebox[\textwidth]{%
\includegraphics[width=1.1\linewidth]{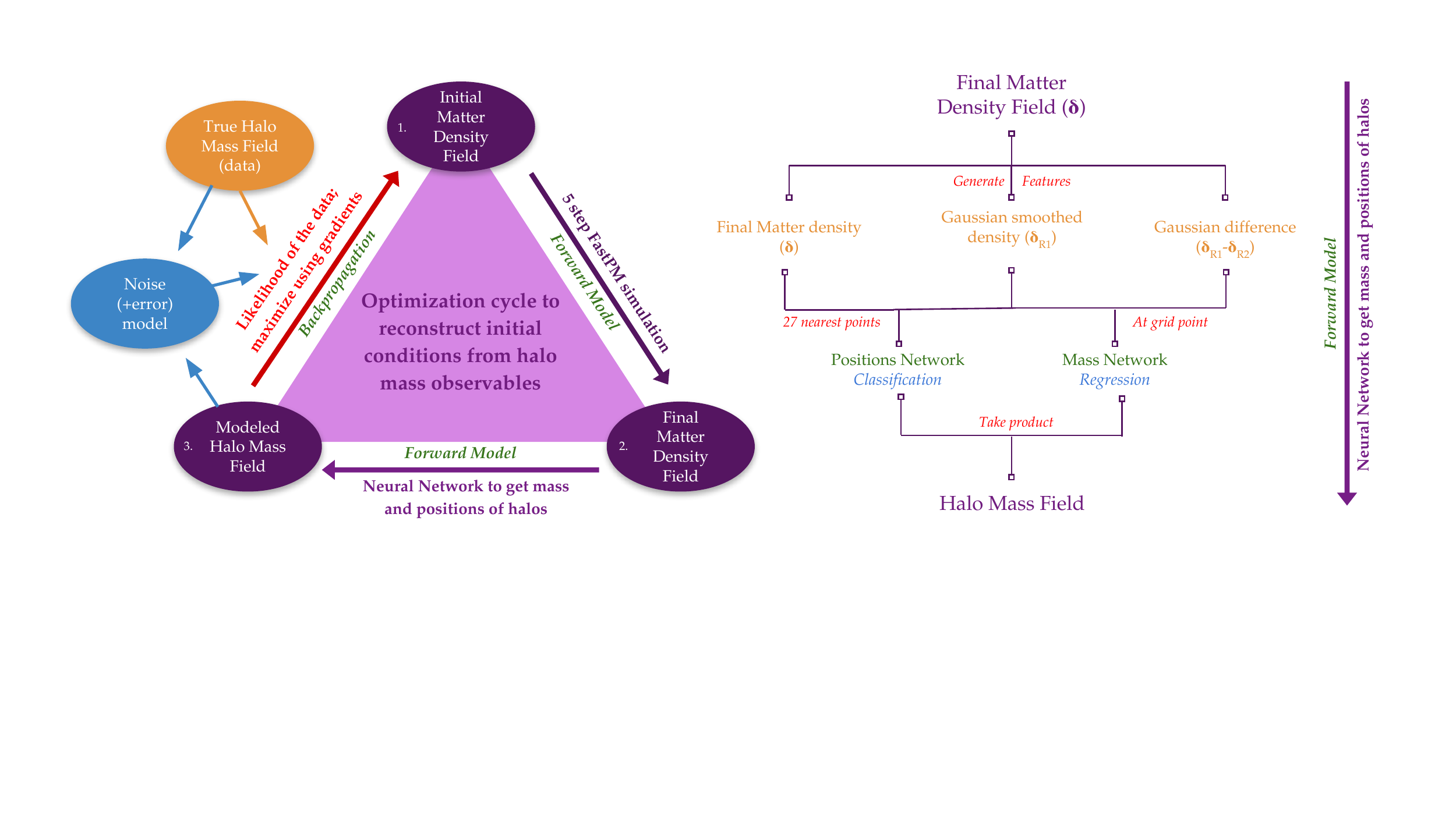}
}
\vspace{-4cm}
\caption{Left: Schematic showing the procedure for reconstruction. Every iteration starts from the initial field, which is evolved to the final matter field and this is then passed through the neural networks to predict the halo mass field. This field is used to estimate the likelihood of the data (Eq. \ref{eq:probnoise}) using a noise model (constructed beforehand from simulations using the trained network, see Section \ref{sec:perfhist}). Corresponding loss function (Eq. \ref{eq:loss}) is then minimized by estimating the gradients with respect to the initial modes and updating them accordingly. Right: Flowchart summarizing the operations involved in second step of the forward model i.e. to predict halo mass field from the final matter field. See Section \ref{sec:model} for details.}
\label{fig:summary}
\end{figure}

The first part of our forward model consists of evolution from linear to nonlinear matter field using FastPM algorithm and reconstruction for this model has already been presented
in \cite{Seljak2017, Feng2018}.
As the next step towards reconstruction from observables, in this work we develop a forward model from the nonlinear matter field to the galaxy observable such as galaxy light or 
stellar mass. We will do so under the simplifying assumption that we have a mean 
relation between the halo mass and the galaxy luminosity (or stellar mass) calibrated using 
observations such as weak lensing \cite{Mandelbaum2006}. This means we can 
work with the halo mass as a proxy for the galaxy luminosity, at least within 
a model with a known scatter between the two. 
To connect the halo masses to the underlying matter density, we propose 
using Artificial neural networks (ANN/NN), which are fully differentiable, 
yet are able to model even very discrete distributions of near pointlike structures.
The schematic for this forward model step is shown in the right panel in \fig{fig:summary}.

Given our proposed forward model, we are able to develop a likelihood function to relate the initial modes and the observed data. This is then combined with the Gaussian prior of the initial modes to construct a loss function that we optimize over to generate a maximum a posteriori (MAP) estimate of the reconstructed initial density field. The schematic for this optimization cycle is presented in the left panel of \fig{fig:summary}.

The motivation of any reconstruction exercise is to be able to extract cosmological information from a tractable statistic. A MAP estimate still leaves us with too many degrees of freedom to extract information in a meaningful way. One needs to be able to estimate the linear power spectrum from these modes because the initial modes are Gaussian and for Gaussian fields, the two point function is the natural summary statistic that captures all of the information. A framework to do so was developed in \cite{Seljak2017}, where one uses simulations to estimate the bias and the mixing of bandpowers in the process of reconstruction and then obtain the correct band powers.
However, for our data, there are other associated forward model and nuisance parameters, such as scatter in the halo mass-luminosity relation. One needs to handle these consistently to be able to marginalize over them and extract the correct underlying cosmological parameters. This is the ultimate goal of our reconstruction, but we will leave this band power reconstruction for future work. Instead, in this work we focus on the reconstruction of the initial field using our forward model and quantify information of reconstruction using 
various two point function statistics.

The structure of the paper is as follows. 
In Section \ref{sec:model}, we describe our model with respect to the architecture and training of the neural networks as well as the features from the non-linear matter field that are used. 
Then in Section \ref{sec:sim}, we describe our simulations and data on which we test our model and reconstruction.
In Section \ref{sec:perf} we evaluate the performance of our model in modeling the halo mass and position field using different metrics.
Next, in Section \ref{sec:recon}, we use our model for reconstruction of initial density modes starting from halos as our data. We develop a loss function, discuss our strategy for optimization and evaluate the performance as well its dependence on the said loss function.
To compare with other methods of reconstruction, we choose standard reconstruction and discuss linear information reconstructed in section \ref{sec:information}.
Finally, we conclude in Section \ref{sec:discussion} along with a discussion of the future work.

\section{Neural Network Model}
\label{sec:model}

\begin{figure}
\centering
\begin{subfigure}{.5\textwidth}
  \centering
  \vspace{2mm}
  \includegraphics[width=\linewidth]{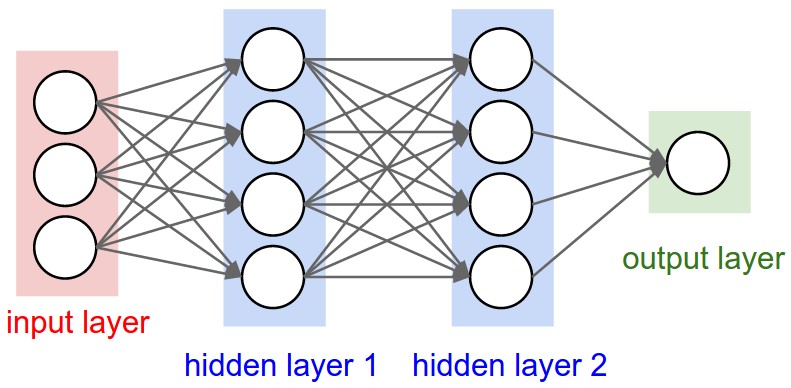}
  \vspace{2mm}
  \caption{}
  \label{fig:nn}
\end{subfigure}%
\begin{subfigure}{0.5\textwidth}
  \centering
  \includegraphics[width=\linewidth]{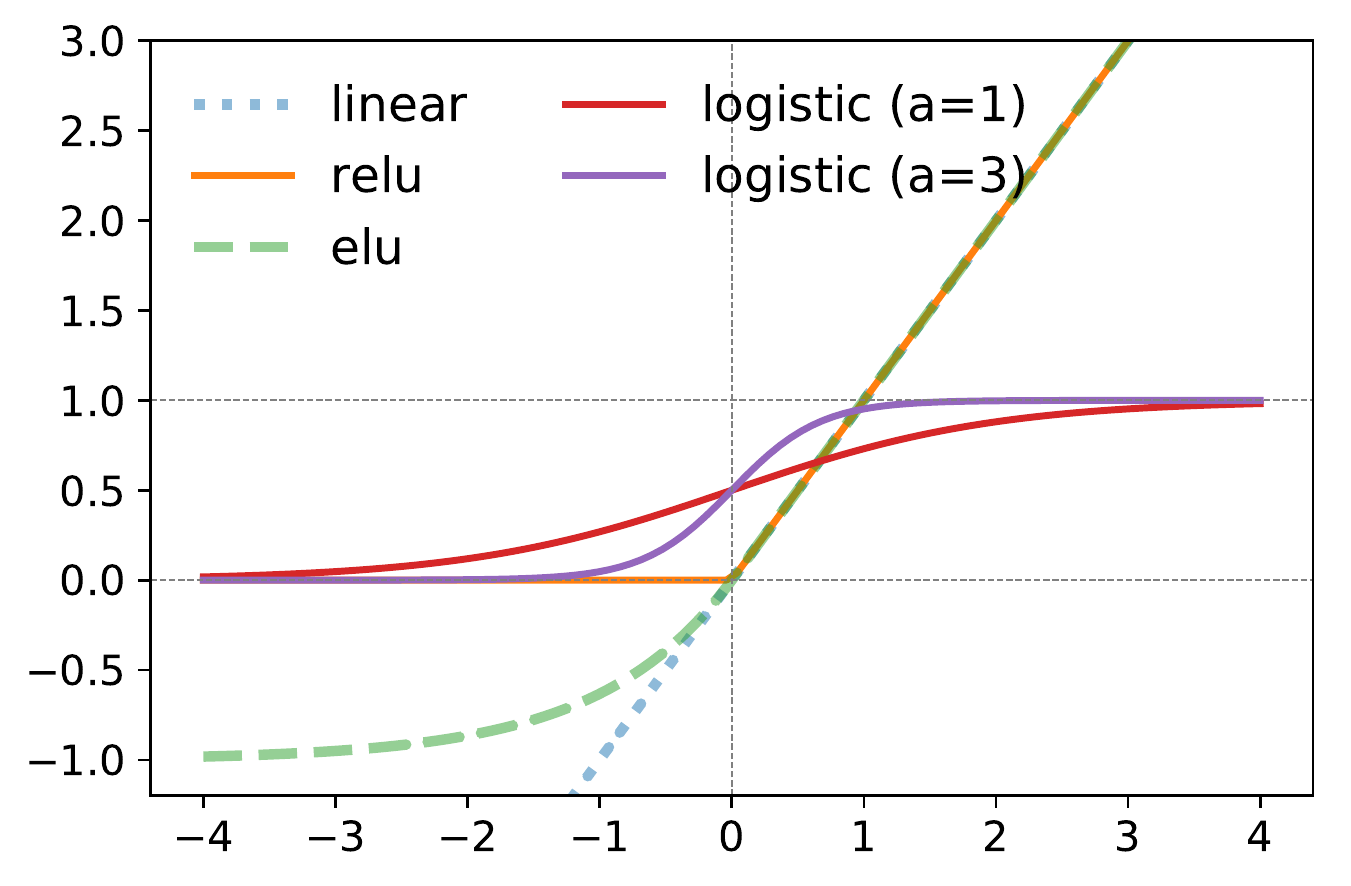}
  \caption{}
  \label{fig:activations}
\end{subfigure}
\caption{(a) A fully connected neural network with 2 hidden layers. Image taken from \url{http://cs231n.github.io/neural-networks-1/}.  \
(b) Different activation functions used in the model}
\label{fig:nnfeatures}
\end{figure}

In this section, we develop our model for predicting the halo mass and position field using artificial neural networks. 
Over last decade, sophisticated versions of these neural networks have come to be widely used in cosmology for diverse purposes, from parameter estimation  \cite{Auld2007}, to model discrimination \cite{Schmelzle2017}, creating virtual universe \cite{Mustafa2017} and to detecting strong lenses \cite{Lanusse2018}. 
While neural networks come in various architectures and sizes, for our purpose here, we use a simple variant called fully-connected neural network. These are directed, weighted graphs where elements called neurons are arranged in layers and each neuron in a layer is connected to all those in the next layer (\fig{fig:nn}), with an associated bias factor (${\bf b}$) and a weight ($W$) and activation function ($\phi$)(\fig{fig:activations}) to determine the output for every directed connection. Thus it recursively takes the input from (${\bf x}$) from previous layers to produce output (${\bf y}$) as
\begin{equation}
{\bf y} = \phi(W\cdot{\bf x} + {\bf b})
\end{equation}
Every neural network requires some underlying features as input to develop a meaningful relationship between these inputs and the data. Given an image, sophisticated convolutional networks (CNN) develop these features themselves by optimizing filters of a restricted size and we could have used them instead of a fully connected network. But in the view of simplicity and understanding, since we have some physical intuition of what information we are seeking and the importance of associated features, we will design the filters ourselves.

The schematic for our model is presented in the form of the flowchart in the right panel of \fig{fig:summary}.
Our model consists of two fully connected neural networks, one to predict the `position' field for the observables and the other to predict the `magnitude' of the corresponding observable. The true observables are the galaxies and their luminosity or stellar mass, but for this first work, we will instead use dark matter halos and halo mass as proxy and model them. Since both the halo mass and galaxy light are continuous properties directly related to the underlying matter content and proportional to each other with some scatter, we expect that it should be straightforward to adapt our model to predict these observables instead of dark matter halo mass. With this in mind, we will henceforth use the term `mass' freely throughout the text to imply dark matter halo mass (for this particular work) and magnitude of any associated observable (in a more general framework). On a similar note, imaging surveys observe galaxies in redshift space. However in this first work as a proof of principle, we work only with halos in real space.

We break down the model into two separate networks for mass and position since it considerably simplifies the problem. Identifying the positions of the observables is now a classification problem which are inherently somewhat easier to solve than regression problems (which we would otherwise be solving to get an observed mass field directly). It is also independent of what the observed quantity corresponding to the mass label is and we can thus use the same trained position network to predict different observables. Identifying the mass after having identified the position of the halos also reduces the dynamic range of regression 
which makes the problem more amenable. 

As mentioned, to make our model differentiable, we have to give up the concept of a pointlike halo and instead we predict the `fields' of these observables. By `fields', we imply that these quantities have been convolved on a grid of ones choosing using a suitable convolution scheme. While such a grid structure is not a fundamental feature of the observables themselves, it arises naturally in any image analysis. Using different grids or convolution schemes should only change the training of our model networks while keeping the conceptual approach unchanged. 
In this work, we consider only CIC convolution scheme, but any other interpolation or smoothing scheme should work as well.

The first network (\textbf{NNp}) identifies the CIC convolved halo positions on the grid. 
Thus, NNp performs classification wherein it identifies if any of the 8 cells (due to the CIC convolution) associated with a grid point hosts a halo and assigns a value between 0 and 1, which is akin to the probability of there being a halo. The output of NNp is thus a discrete, position mask for the halo field (see \fig{fig:pmask}).
The second network (\textbf{NNm}) on the other hand performs a regression on the underlying features to predict the value of CIC convolved mass at a given grid point, assuming that there is a halo associated with that grid point. This leads to a continuous mass field (see \fig{fig:mmask}).

Given these two networks, our model for the discrete halo mass field is simply the product of these two fields.
$$
\rm{model = NNp \times NNm}
$$
For the sake of brevity and to avoid repetition, henceforth 
we will say that neural network predict the halo position and the halo mass 
when we actually mean that it predicts the CIC convolved position field and CIC convolved mass field. 
We will also use NNp and NNm to refer to both, the networks as well as their outputs
when it will not lead to any confusion.

\subsection{Architecture}
\label{sec:arch}
As described above, fully connected neural networks consist of neurons arranged in layers 
with different activation functions to introduce non-linearities in the model. 
The choice for the number of layers, number of neurons in every layer and the activation
functions are the hyperparameters that need to be tuned. We have explored different values of these hyperparameters and  settled on the following architecture for the two networks:
\begin{itemize}
\item Position network (NNp): This network consists of 2 hidden layers with 50 and 30 neurons in the hidden layers, followed by an output layer of size 1. 
The activation functions used for the neurons in the hidden layers is ReLu (Eq. \ref{eq:relu}) and the output layer neuron is followed with a logistic function to squash the output value between 0 and 1. We also sharpen the logistic function by multiplying the exponent with ad-hoc factor of $a=3$ (Eq. \ref{eq:logistic}) to increase the discreteness of the output.
\item Mass network (NNm):  This network consists of 2 layers with 20 and 10 neurons in the hidden layers with ELU activation functions and a linear activation function followed by the final layer which is one neuron in size. 
We find that using ReLu activation instead of ELU (Eq. \ref{eq:elu}) also works well and but the latter is easier to train.
Moreover, it turns out that a simple quadratic regression  performed over the same underlying features (features described in the next subsection) also performs comparably to the neural network for some grid resolutions and halo number densities. However overall, the networks give more robust performance across different configurations.
\begin{eqnarray}
\label{eq:relu}
{\rm {ReLu}}(x) &=& \begin{cases}
  x & \text{if $x>0$}\\
  0 & \text{if $x\leq$0}
  \end{cases}\\
\label{eq:elu}
{\rm{ELU}}(x) &=& \begin{cases}
  x & \text{if $x>0$}\\
  e^{x}-1 & \text{if $x\leq0$}
  \end{cases}\\
\label{eq:logistic}
{\rm{logistic}}(x, a) &=& \frac{1}{1 + e^{- a x}}
\end{eqnarray}

\end{itemize}

In general, the performance of the models seems to be somewhat insensitive to the number of neurons beyond a given size. For example, we find that NNp gives similar performance with (30, 20) neurons in the hidden layers, but we choose (50, 30) architecture to be more conservative. 
The performance of networks, especially NNp, improves significantly on going from single hidden layer to 2 hidden layers, while a third hidden layer improves things marginally. Traditionally, practitioners of fully connected networks avoid going deeper than two or three layers since deeper networks are harder to train (this 
changes for deep CNN type networks where a third dimension of filter shapes is added). For our purpose, we settle with  a 2 layer fully connected architecture in this work because we find satisfactory performance with this simplistic architecture. We also anticipate that more complex architectures will introduce difficulties taking gradients of forward model due to the possibility of vanishing gradients (due to activation functions saturating intermediate outputs). It is still likely that sophisticated architectures such as convolutional neural networks (CNNs) will improve the performance of our models and open doors to further applications such as generating mock catalogs of halos from density fields. We leave such forays for the future.

\subsection{Features}
\label{sec:features}
The neural network requires a set of underlying features (ideally physically motivated) 
as inputs. These features should provide meaningful information to learn their relationship with the corresponding observed data and hence to be able to predict the latter from the former. 
Since we are interested in predicting the halo mass field, and all halo formation models predict that more halos form in regions of higher overdensity, the underlying final matter density field provides a natural candidate for these features. 
The simplest models such as Spherical collapse \cite{Gunn1972} simply predict that halos form where
overdensity exceeds a certain threshold with the mass of the halos corresponding to the largest scale at which it does so \cite{Press1974,Bond1991}. 
Other halo formation models such as Ellipsoidal collapse advocate the role of shear \cite{Sheth1999, Sheth2001} while assembly bias models suggest that other environmental features \cite{Han2018,Mao2018} also play a role in halo formation.
In principle, one can use all these fields as underlying features and this should ideally improve the accuracy of predictions, albeit the gains gradually diminish \cite{Lucie-Smith2018}. The increasing complexity of the model will also lead to increased difficulty of training the neural networks.

Given the simplicity of our neural network and the approximate nature of our simulations,
we use only the following three fields to generate the feature vector - 
\begin{itemize}
\item CIC convolved matter density field ($\delta_0$)
\item Density field smoothed with a finite Gaussian kernel on some scale ($\delta_{R1}$) 
\item Difference of finite Gaussian smoothed fields on two different scales ($\delta_{R1} - \delta_{R2}$)
\end{itemize}

\begin{figure}
\centering
\begin{subfigure}{.45\textwidth}
  \centering
  \vspace{2mm}
  \includegraphics[width=\linewidth]{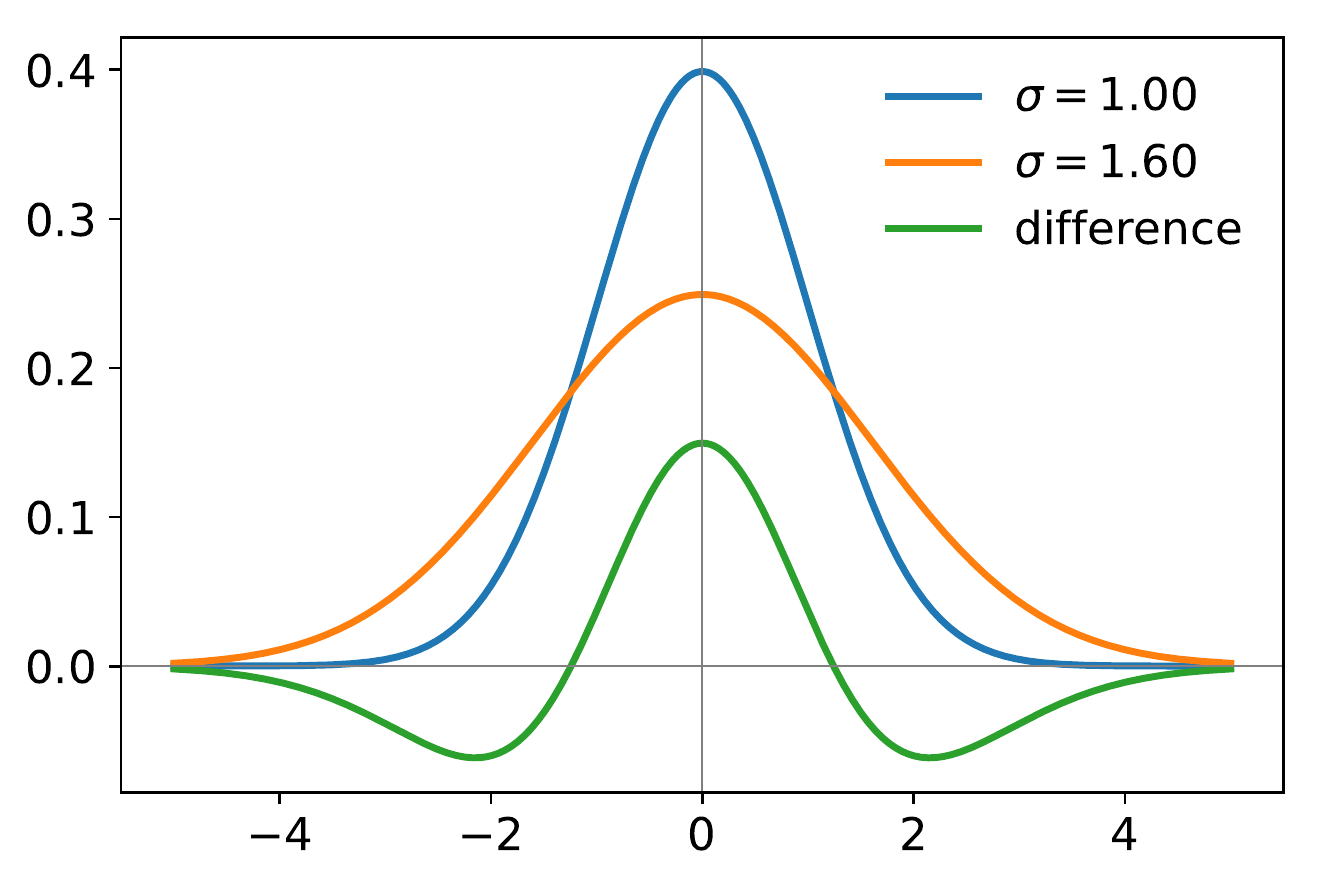}
  \vspace{2mm}
  \caption{}
  \label{fig:gd}
\end{subfigure}%
\begin{subfigure}{0.55\textwidth}
  \centering
  \includegraphics[width=\linewidth]{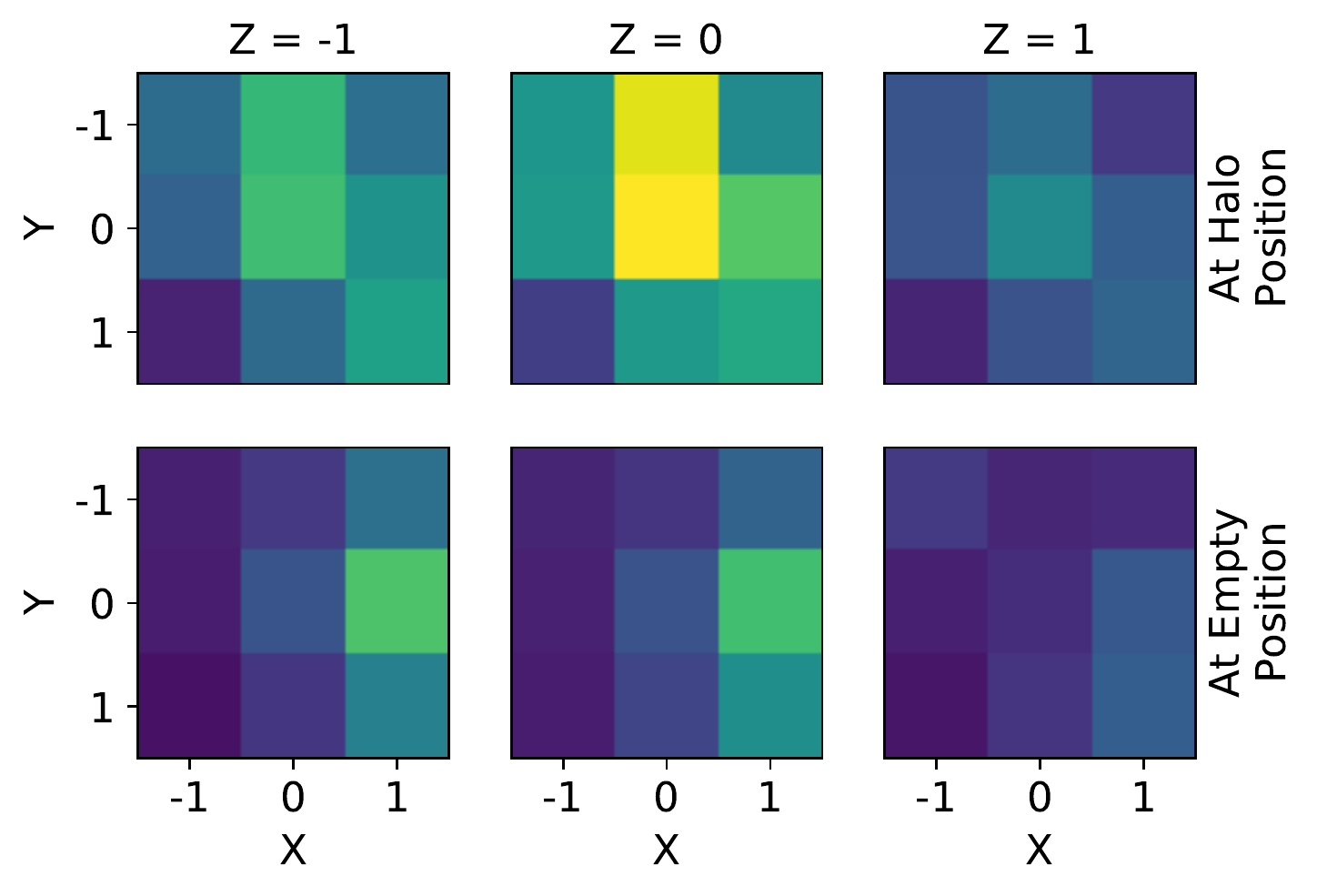}
  \caption{}
  \label{fig:27ft}
\end{subfigure}
\caption{(a) Difference of two Gaussian smoothed fields (green) localizes the peaks better than either of the fields.
(b) The density field at 27 nearest neighbor points of the central grid point (X=0, Y=0, Z=0; Z-coordinate in the title is increasing going from left to right) for 2 cases - (top row) when the central grid point corresponds to a halo position, (bottom row) central grid point is an empty grid point in the field.
In the top row, local maxima is visible around the halo- along X axis (middle row in Z=0 square), Y axis (middle column in Z=0 square) and Z axis (central cells in the three squares). In comparison, the bottom panel
looks more like a smooth gradient.}
\label{fig:features}
\end{figure}

We do not use other aforementioned features such as shear and velocity field because to speed up reconstruction, we work with approximate simulations on coarse resolutions and small number of time steps. For these simulations, we find that these features do not significantly improve the performance of the networks. Instead, we use density fields smoothed at three different scales (mesh scale, $R_1$, $R_2$) as an approximation to environmental information supplementing local density information. 
In addition, we use Gaussian difference (GD) which is an approximate blob detection technique 
used in computer vision \cite{Lowe2004} to identify peaks in this space.  
Since halos are more likely to form at peaks in density fields, we find that 
GD localizes the position of halos by identifying peaks in high density blobs which are likely to cross a simple density threshold over more than one grid points for a single halo (\fig{fig:gd}). 

Halo formation is a local process and hence we need only local information to predict halo masses and positions with our model. Thus, the feature vector for the mass network (NNm) consists of the values of aforementioned three fields at the grid point of interest and has dimension 3.
However for the feature vector corresponding to a grid point in the position network (NNp), we use the values of the three fields from all the 27 points in a ($3 \times 3 \times 3$) cube around that point. 
This is motivated by the fact that the preferred locations of halo formations are density peaks above some threshold and given  three consecutive points in any dimension, it is easier to  identify the positions of these density maximas (see the caption of \fig{fig:27ft} for description). Indeed, this improves the performance significantly for the position network over using the values only at the grid point, as is done for the mass network. 
This is also similar to what a CNN does, wherein its every neuron identifies the filters by convolving spatial pixels to maximize the information. The gains for NNp with the convolved filters suggests it would be worth exploring further whether using a full CNN architecture with multiple filters could improve the learning. For now, the dimension of the feature vector of our position network is $3\times 27=81$.

\subsection{Hyperparameters}
\label{sec:hyperparams}
Hyperparameters are parameters that are not learned during the training of the neural networks. Model parameters such as weights and bias are optimized during training, while other parameters like number of hidden layers, number of neutrons etc. are in practice set by hand and then configured for best performance by simply exploring the parameter space.
For both networks, we fix the architecture after a couple of trials and hence do not change the associated hyperparameters such as number of hidden layers, number of neurons, activation function etc. Amongst parameters associated with training, we find that the performance is most impacted by varying batch size (the number of data points used in training at every iteration before updating weights with back-propagation) and regularization strength (penalty imposed on the norm of weights and biases to avoid over-fitting) and hence we have performed a grid search over these to get optimal values. 

In addition, we also introduce other heuristic parameters to generate the training data set and to assist in training. The straightforward way to generate a training data set would be to simply use the full simulation. However, since the halos occupy a small fraction ($\sim 5\%$) of the grid points, using the whole simulation leads to severely imbalanced class ratio and it becomes much harder to train the networks. We also do not use the traditional ways of re-weighting and oversampling \cite{He2009} to address this imbalance since most, \emph{but not all}, high density regions host halos and using both these methods skews the ratio of halo to non-halo positions in these regimes. We find that training this way indeed leads to overly connected structures in the regions of high overdensity. 

Thus, we introduce other parameters to leverage our knowledge of halo formation and  construct better representative samples to balance the classes by reducing noise and redundancy in data.
For training the position network to perform classification, instead of using the whole simulation, our training dataset consists of all the positions with halos to create one class (1) and then we sub-select certain number of empty, non-halo positions to fill the other class (0). 
Based on our domain knowledge, we divide these empty positions into 4 types and keep the number (fraction) of each type as a heuristic parameter. These different kinds of  empty positions are - 
i) grid points that have a smaller halo below the abundance threshold of halos,
ii) grid points above a heuristically chosen matter overdensity threshold, 
iii) grid points below the density threshold 
(we find that this is quite degenerate with the grid points having smaller halos in (i) and hence we end up not varying the number of these points as a parameter). 
In addition, we iv) keep a limiting value parameter to limit the CIC convolved value on the grid above which we accept the presence of halo. Therefore we do not choose all the 8 grid points to which CIC scheme interpolates over. 
Similarly for NNm, instead of using the full simulation, we use a parameter which selects only the points assigned a value above a threshold by the position network (NNp).
This latter parameter is to guard against bad fits driven by points which are not going to be selected by the position network anyways. 
Thus upon including the batch size and regularization strength, we have 5 hyperparameters for NNp and 3 for NNm. 

\subsection{Training}
\label{sec:train}
We use \textit{Adam} \cite{Kingma2014} as our optimization algorithm during training. We gauge the performance of our networks by using L1 loss function for the position network and 
L2 loss function for the mass network. To prevent over-fitting, we use L2 (ridge norm) for regularization in both networks as well as dropout of neurons with drop-probability of 0.3.
We perform training on NERSC using \textit{Keras}\footnote{\url{https://github.com/keras-team/keras}} with \textit{Tensorflow}\footnote{\url{https://github.com/tensorflow/tensorflow}} backend \cite{Chollet2015,Abadi2016}. 
To generate the training data set, we combine data from three different simulations using the hyperparameters described in the last subsection and use a fourth simulation as validation set. The validation set is not used for training but we monitor the accuracy (loss) on this validation set over iterations and use it as the early stopping criterion, i.e. when the change in accuracy (loss) falls below 
a certain threshold value consecutively for a given number of epochs, we stop training further. 
Depending on the batch size, it took between 1-5 minutes to train a network on a single core, 2.3 GHz Intel Xeon Processor with 512 GB memory, on Cori supercomputer at NERSC.
To tune hyperparameters, we perform a simple grid search over different parameters, 
and it takes 8-15 hours get a network trained for all the parameter combinations. We finally choose the networks that performed the best at the level of the two point functions (cross-correlation and transfer function, as described in Section \ref{sec:perf2pt}) as our model for reconstruction.
For the purpose of this work, we trained a different network for every spatial resolution and number density of halos, but in principle one can train across different resolutions. We will explore this in the future.  

\section{Simulations}
\label{sec:sim}

We use the halo catalog and density fields simulated using 
\texttt{FastPM}\footnote{\url{https://github.com/rainwoodman/fastpm}} code \citep{Feng2016}. 
\texttt{FastPM} is a PM simulation to generate non-linear dark matter and halo fields in a fast manner, employing much fewer time steps and enforcing linear Zeldovich 
displacement on large scales to reproduce the results of full N-body simulation. 
Despite its approximate nature, in \cite{Feng2016} it was shown that the code performs extremely 
well on various benchmarks such as the dark matter power spectrum, the halo mass function and the halo power spectrum. 

The simulations have a flat $\Lambda$CDM cosmology model with the Hubble parameter $h =0.6711$, 
matter density $\Omega_M = 0.3175$, baryon density $\Omega_b = 0.049$ and spectral index $n_s = 0.9624$. 
The linear power spectrum used was generated with \texttt{CAMB}.
The initial conditions of particle displacement and velocity were computed at second order in 
Lagrangian Perturbation Theory at redshift $z=9$.
The underlying density fields which are used as features for the neural networks are generated using 
5 step \texttt{FastPM} simulations since this is also the forward model that is used during the optimization.
To generate the data, the halos are identified in a 4x higher resolution \texttt{FastPM} simulations with 40 time steps, starting from the matched initial conditions.
Halos are identified using FOF halo finder in  \texttt{NbodyKit}\footnote{\url{https://github.com/bccp/nbodykit}}
\cite{Hand2017} with the optimization presented in \cite{Feng2016a} using a linking length of 0.2.
We use smaller number of timesteps for the features and optimization to reduce the cost of reconstruction which scales linearly with the number of timesteps. All the results are presented for periodic cubic box of size $L=400$ Mpc/h and $N=128^3$ mesh (the corresponding FOF halos were identified using $N=512^3$ simulation) and number density of halos $\bar{n} = 10^{-3}{\rm (h/Mpc)^3}$ unless otherwise specified. 
We have verified that we get similar performance for other box sizes, resolutions and number densities.

\section{Model Performance}
\label{sec:perf}

In this section, before using our NN model for reconstruction, we evaluate the performance of the model
in modeling the halo position and mass fields of FOF halos.
In section \ref{sec:perfvis}, we present the visual output of the two networks and qualitatively compare
this with true halo fields as well as quantitatively discuss the performance in predicting the position of halos. In section \ref{sec:perfhist}, we estimate the error made in predicting the halo masses and this forms
the basis of our loss function which will be optimized over in Section \ref{sec:recon}. 
Lastly, in section \ref{sec:perf2pt}, we gauge our model at the level of two-point functions of interest. 
 
\subsection{Visual Comparison}
\label{sec:perfvis}

\begin{figure}
\centering
\begin{subfigure}{.5\textwidth}
  \centering
  \includegraphics[width=\linewidth]{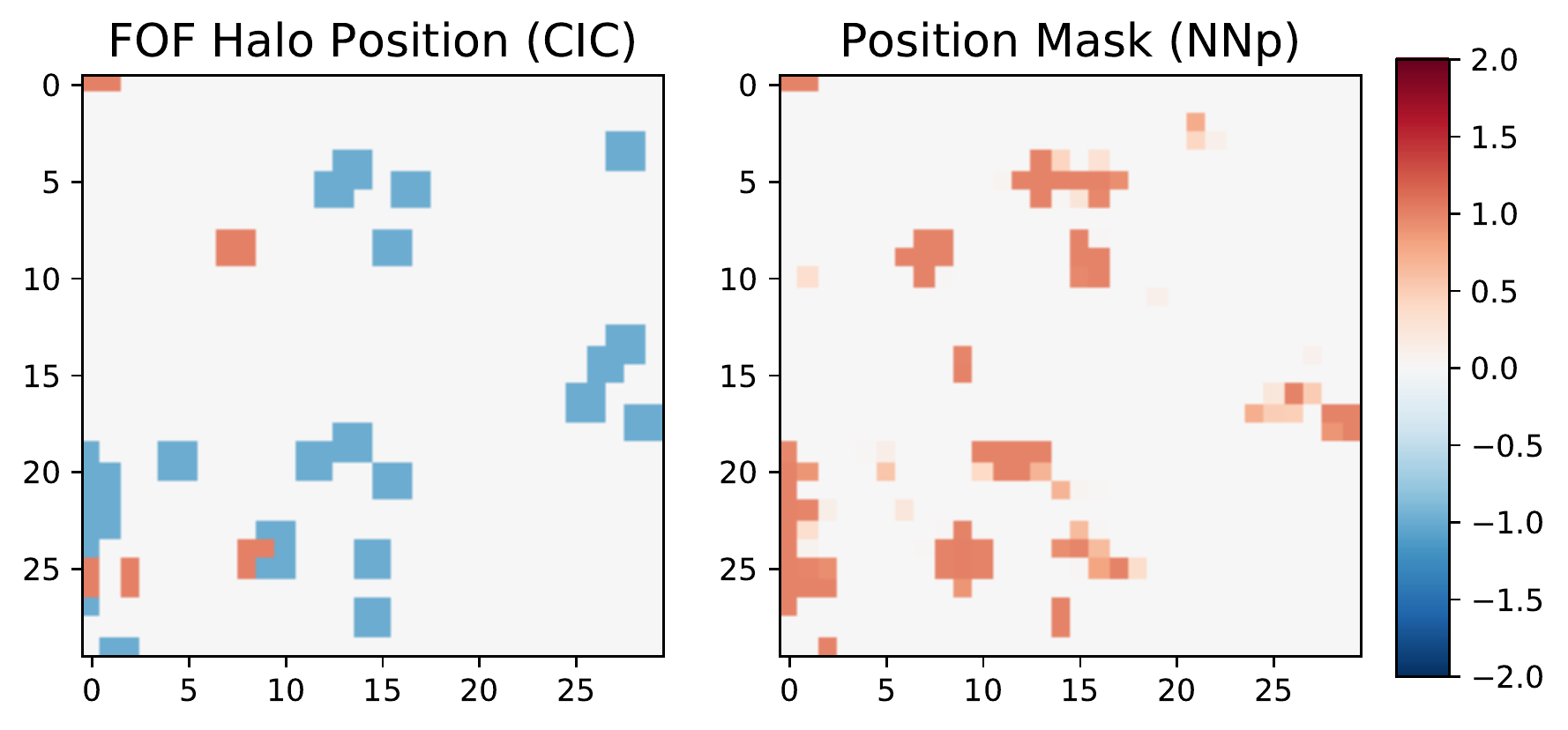}
  \caption{}
  \label{fig:pmask}
\end{subfigure}%
\begin{subfigure}{0.5\textwidth}
  \centering
  \includegraphics[width=\linewidth]{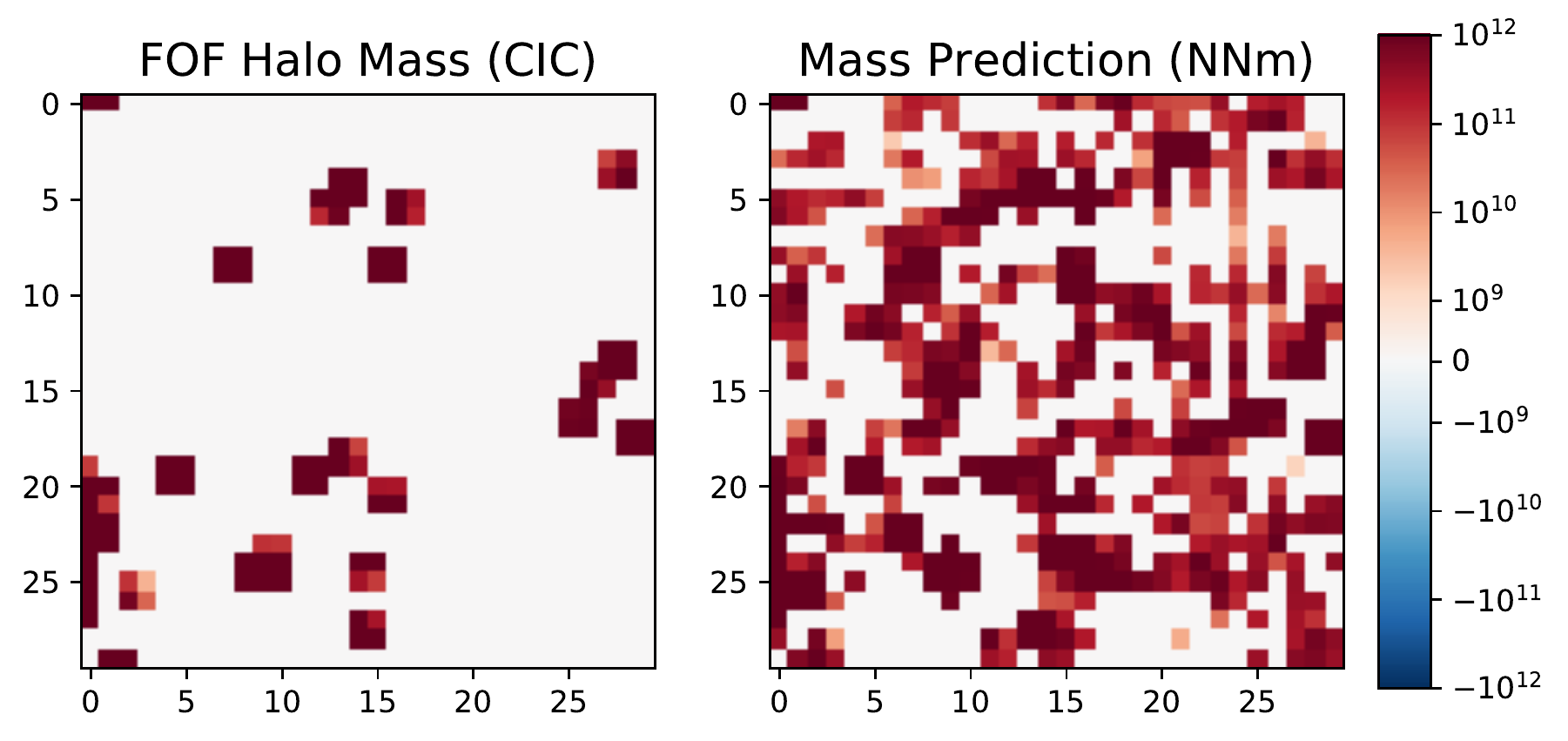}
  \caption{}
  \label{fig:mmask}
\end{subfigure}
\caption{(a) Comparing output of NNp, position mask with the FOF halos. Since halos are convolved with CIC, they appear as squares in the projection. In the FOF halo position (left figure), we have distinguished between the massive halos found in $128^3$ simulation (which provides all the density field features for the neural network) in red, and the smaller halos supplemented from $512^3$, shown in blue, to reach the requisite number density. 
(b) Comparison of regression output of NNm with FOF halos, without imposing the NNp mask on the prediction of NNm}
\label{fig:masks}
\end{figure}

\begin{figure}
\centering
\includegraphics[width=0.5\linewidth]{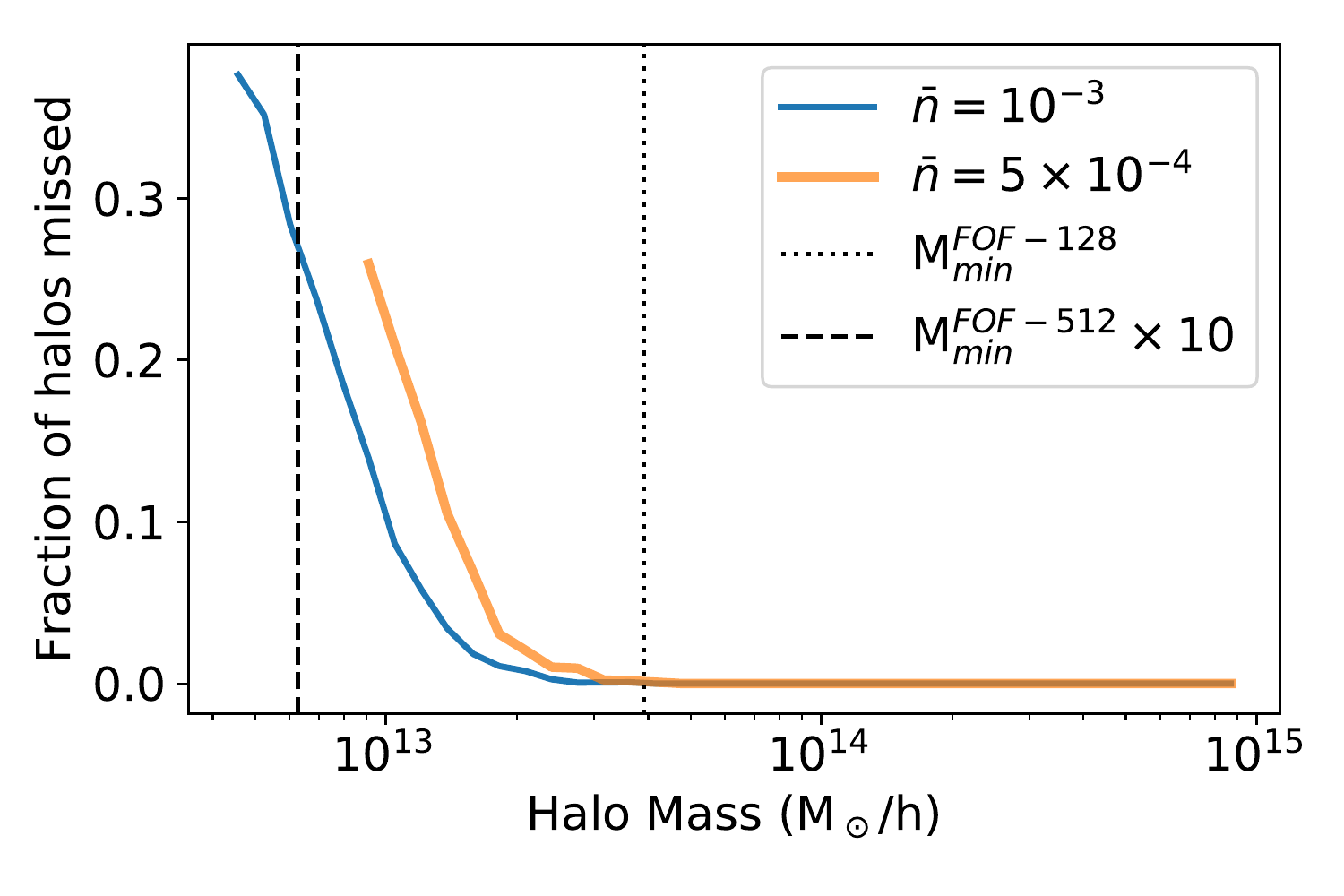}
\caption{Fraction of halos (Nearest-neighbor points) missed by the position mask created by NNp network as a function of mass. For reference, the dashed and the dotted lines show the mass of the smallest halo with 12 particles identified in the simulation when run with $128^3$ grid- which generates the density features; and $512^3$ grid- which generates the data. Neural networks are able to identify the halos with mass upto $\sim$ 8 times below the FOF threshold.}
\label{fig:missed}
\end{figure}

We compare the output of the neural networks, NNp and NNm, with the FOF halos in \fig{fig:masks}.
We show a local snapshot of single slice for all the four grids.
Since the halos have been CIC interpolated to the nearest 8 grid cells, they are squares in the 2D projection. 
In the FOF slice in \fig{fig:pmask}, the red points correspond to halos identified in the low-resolution simulation ($128^3$) while the blue squares are smaller halos identified in a 4x higher resolution simulation ($512^3$)
to reach the requisite abundance of $\bar{n}= 10^{-3}{\rm (h/Mpc)^3}$ in $400$ Mpc/h box. 
Thus \fig{fig:pmask} shows that our neural network is able to identify the halos going much lower in mass ($\sim 8$x in this configuration, see also \fig{fig:missed}) than the FOF threshold of the simulation, set by the particle mass. This motivates the application of neural networks to generate halo catalogs for future surveys using coarse PM simulations. 

\begin{figure}
\centering
\begin{subfigure}{.5\textwidth}
  \centering
  \includegraphics[width=\linewidth]{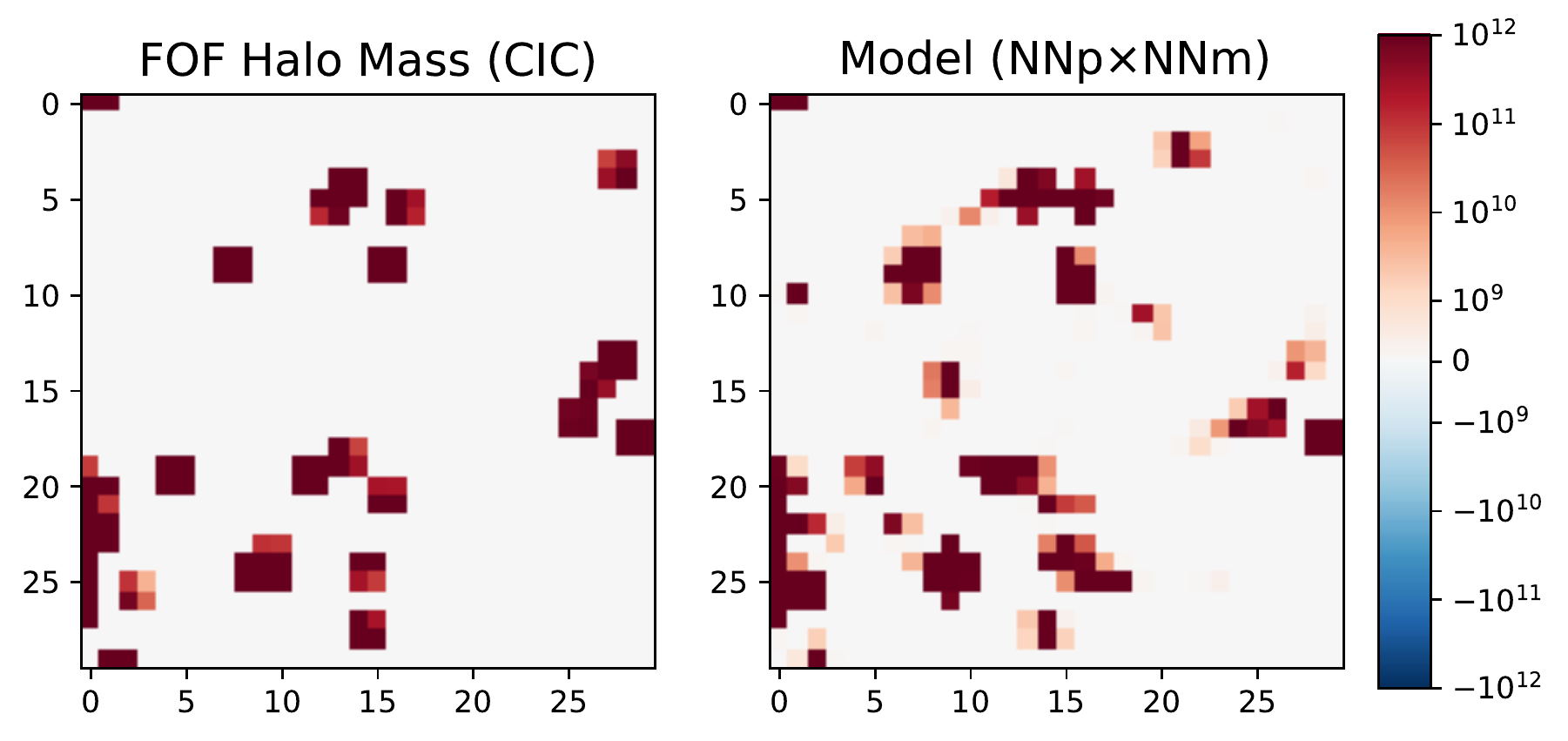}
  \caption{}
  \label{fig:model}
\end{subfigure}%
\begin{subfigure}{0.5\textwidth}
  \centering
  \includegraphics[width=\linewidth]{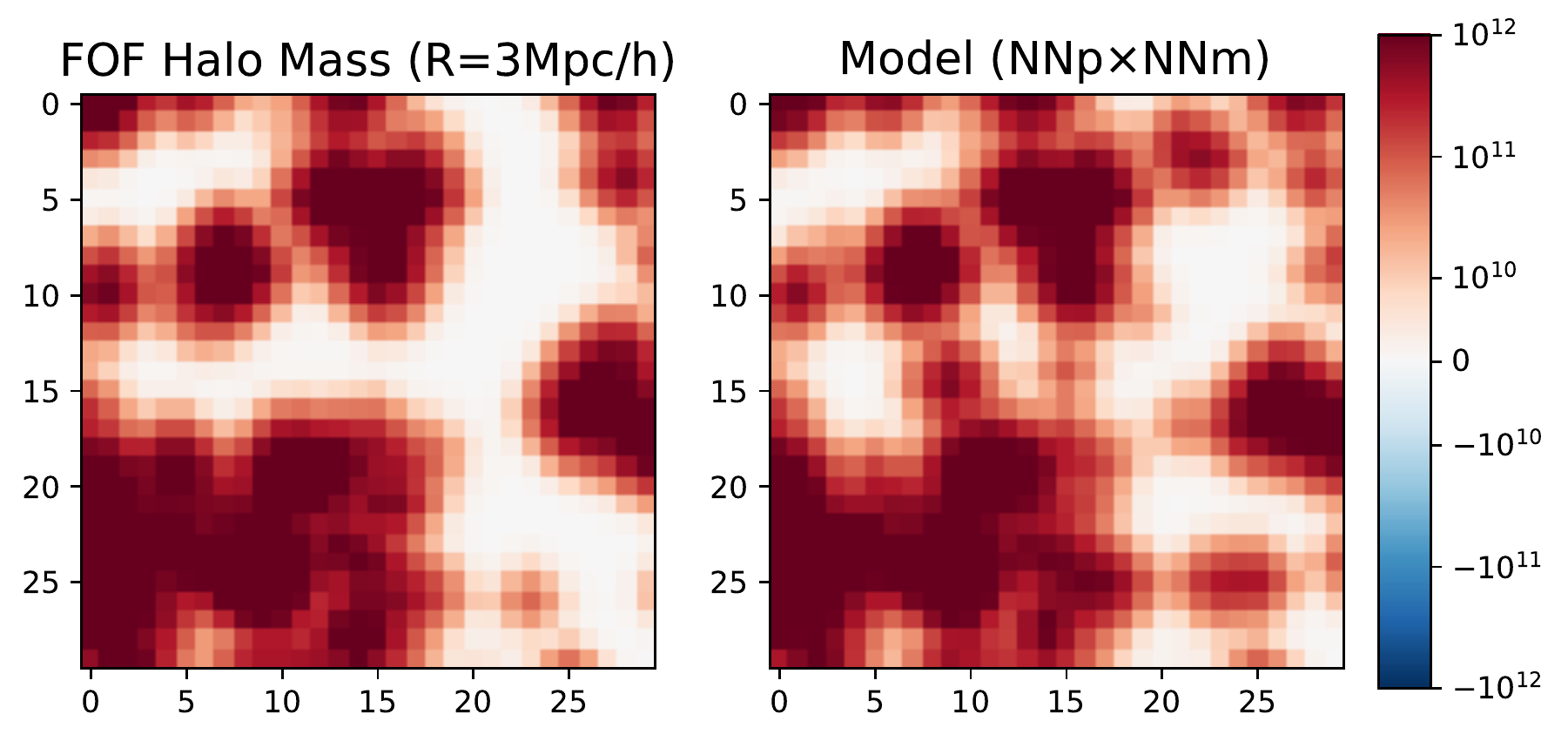}
  \caption{}
  \label{fig:modelR}
\end{subfigure}
\caption{(a) FOF halo mass field and the model prediction (NNp $\times$ NNm)
(b) Same as (a) but smoothed with a Gaussian kernel of $3\mpch$.}
\label{fig:models}
\end{figure}

The output of the mass network, NNm, is shown in \fig{fig:mmask}. Since this network performs a simple regression, 
there are huge connected areas in regions of high density. To convert this to a discrete field corresponding to 
the halo mass field, this is multiplied with NNp. This is shown in \fig{fig:model}, which has some more non-zero
grid points than \fig{fig:pmask}. This is because the position mask of NNp, is not a binary mask of 0s and 1s but rather
a continuous output of logistic function (Eq. \ref{eq:logistic}) and hence some high density (and correspondingly high mass) positions of NNm are only suppressed while not being quite set to zero. This continuity of logistic function
is needed so we can take the gradients: in the limit of very high resolution the halo positions can be very precise, but if the field is discretized into Dirac delta functions with the use of a binary step function,  
the gradient will not be able to tell the field which way to move to reconstruct these delta functions. In the opposite limit of 
a very high smoothing (width of logistic function, set by the ad-hoc parameter $`a'$ in Eq. \ref{eq:logistic}) one looses small scale information in the presence of the noise. 
The choice of width scale, $a=3$ (\fig{fig:activations}), is thus a compromise between these different requirements.
We note that for the choices of 
voxel scale and number density of halos made here most of the voxels are empty (\fig{fig:masks}), 
which indicates that we have a high spatial resolution and we should be more concerned about singular gradients than loss of information due to smoothing. 

To quantify how well we do in predicting positions of halos, we look at the two kinds of errors that 
the position network makes - it misses some halos; while it also identifies some extra points as halos which 
are not present on the FOF grid. To quantify these errors, we can measure the `Missed fraction' and `Empty Fraction'.

\begin{eqnarray}
\rm{Missed\ Fraction} &=& \frac{\#(\rm{NNp}_{<0.5}\ \cap\ \rm{CIC}_{>0})}{\#\rm{CIC}_{>0}} \\
\rm{Empty\ Fraction} &=& \frac{\#(\rm{NNp}_{>0.5}\ \cap\ \rm{CIC}_{=0})}{\#\rm{CIC}_{>0}}
\end{eqnarray}
where NNp(CIC) are the values at the grid points on the NNp(CIC convolved halo position) grid and 
$\rm{NNp}_{<0.5}$ for e.g. denotes the points where the said value is less than 0.5.

Missed fraction quantifies the number of halo locations that the neural network does not detect with 
the probability threshold of $0.5$. For the number density
of $\bar{n} = 10^{-3}\ (5 \times 10^{-4}){\rm (h/Mpc)^3}$, when compared to CIC positions of halos (thus assigning 8 grid
points to individual halo), missed fraction is around 
$\sim 25\%\ (15\%)$ while it drops to $\sim 15\%\ (7\%)$ when compared to the nearest neighbor gridding of halos.
Since the latter assigns a single point per halo, it quantifies the actual number of halos missed completely by the model.
Mass distribution of the missed halos as a fraction of the total number of halos 
(halo mass function) is shown in Figure \ref{fig:missed}. As expected, we find that the model accurately detects high mass halos and the missed halos are primarily of lower mass. We note that at these masses, finding the correct halos is hard, since most of 
the halos are close to the mass threshold, and whether or not a halo should be above or below the set threshold will 
thus depend on small details. This problem is somewhat alleviated if one is looking at the halo field weighted by the 
halos mass. This is explored further below. 
 
Empty fraction corresponds to false positives and 
quantifies the extra points that the network predicts as halo positions but which are not in the FOF catalog. 
For CIC gridding, this is $\sim 10\%\ (30\%)$ for the two number densities. However, most of these positions are still adjacent to halos, 
and if we do not include the points which are within 1 cell of any halo, 
the empty fraction drops to $\sim 1\%\ (5\%)$.
The grid points in this category which are falsely detected as halo positions are primarily of 2 types - 
locations where center of FOF halos and density peaks positions are widely separated from each other, 
and locations which have lower mass halos and hence are below the abundance cut imposed on the catalog. 

\subsection{Error Histograms}
\label{sec:perfhist}
Having a forward model that predicts the halo mass or some other property is not all that is needed: to do reconstruction via optimization, one also needs a likelihood model for the data, i.e. an error/noise model which can be combined with the prior to construct a loss function. This error model should be sparse, so that it can 
be efficiently evaluated even on large grids. 
We are thus interested in estimating the error made by the model (NNp$\times$NNm) in predicting mass
over all points on the grid to create our noise model. However from the previous section, we know that NNp and hence the model
has false positives, most of which lie close to a halo. 
If we directly calculate the error at the level of the two fields as is, 
we will be ignoring this information that we are in the correct neighborhood, if not at the exact position.
This information has the potential of helping with the convergence of the optimizer during the reconstruction. 
To make use of this local information, we calculate the error in mass estimation
on a smoothed model and data field instead. This smoothing tells the gradients which way to move to construct the peaks corresponding to the halos. 
We thus smooth both the modeled and the halo mass field with Gaussian smoothing kernel of $3 \mpch$ (since that was also the minimum scale used to generate the input features for the neural networks).
These smoothed fields are shown in \fig{fig:modelR} and they are in better visual agreement than the discrete fields (\fig{fig:model}).

\begin{figure}
\centering
\includegraphics[width=\linewidth]{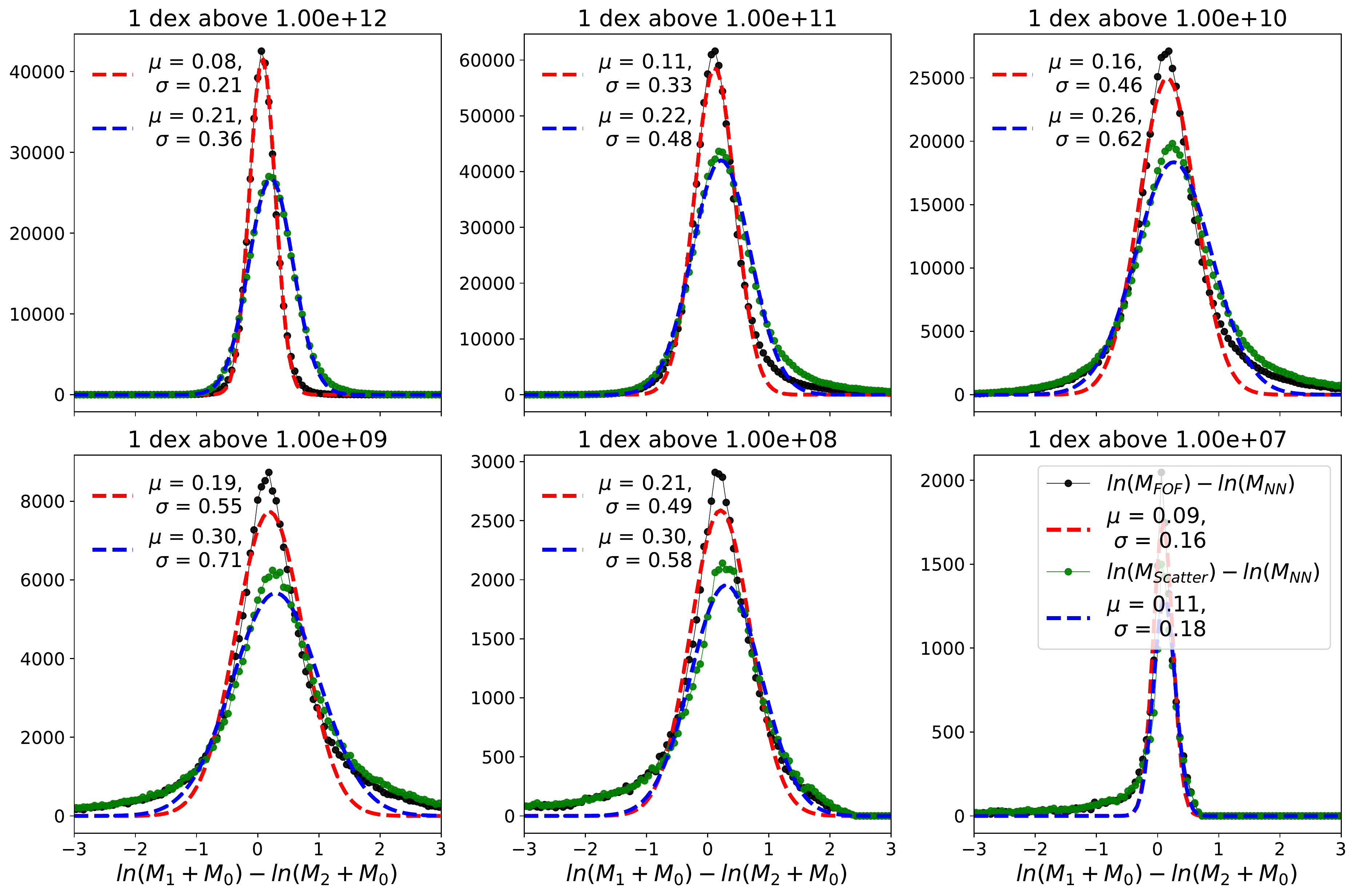}
\caption{Modeling error histograms. Error is the difference between the smoothed mass fields of the data (FOF halos) and neural network predictions. It is estimated in terms of $\log(M_R + M_0)$ where $M_R$ is the value of the smoothed halo mass fields and $M_0$ is a constant nuisance parameter, here $M_0 = 10^8 M_\odot$/h. Every panel corresponds to different positions (bins) on the data grid where the data (mass) value at those points is in the range given by the titles of the panels. Black points are the data points from the simulation assuming the correct mass of the halos ($M_1 = M_{\rm FOF}, M_2 = M_{\rm NN}$), green points are with log-normal scatter of 0.2 dex in halo masses for the data ($M_1 = M_{\rm scatter}, M_2 = M_{\rm NN}$). The dashed lines are the log-normal best fits with mean and standard deviation specified in legend.}
\label{fig:noise}
\end{figure}

We then compare these two smooth fields at point by point basis to ascertain the error in our model. 
This is shown in \fig{fig:noise}, where we show the histogram of the error made by the model in predicting logarithm of the mass at different points
binned as a function of the true mass at those points (different panels). We also want an analytic form for the error PDF.
We find that displaced log-normal model is effective for this purpose, so 
we calculate the difference in $\log (M_R + M_0)$ with $M_R$ being the value of the smoothed mass fields and $M_0$ a nuisance parameter, 
its value being $10^8 M_\odot$/h in \fig{fig:noise}.
We bin grid points based on the value of $M_R$ for the FOF halo mass fields in different panels as specified by the title and plot the corresponding error histogram.

The black points in \fig{fig:noise} are the modeling error data points, measured against true FOF halo masses
convolved on the grid which were also used for training NNm network. 
The red dashed plots are the log-normal fits to the error histogram which fit the peak well but underestimate
the tails. There is a non-zero offset for all mass bins, and it is taken out by our model.  
The error (standard deviation) increases as we go lower in mass until $M_0$ starts to dominate. 
Thus, as in position (\fig{fig:missed}), our model predicts masses of heavier halos better than lower mass halos.

So far, we have assumed that we know the true masses of the dark matter halos.
However this is not true in actual surveys. LSS surveys observe galaxies and there is an intrinsic
log-normal scatter associated with stellar mass, luminosity or other observables and the corresponding halo mass \cite{Rodriguez-Torres2016}. If we constrain the halo masses by weak lensing, there is still an error associated with this calibration. To estimate our modeling error in the case of noisy halo masses, we add a log-normal scatter of 0.2 dex before matching abundance of halos and make the error histograms for our model against these noisy halo masses. 
These are the green points in \fig{fig:noise}. The blue plot is the log-normal fit to this modeling error. 
The error (standard deviation) for the scattered data is larger than for the unscattered data i.e. when we assume we know the correct halo mass. Further, the offset has also increased over the true cases. 
However we are still able to fit the error histograms with a displaced log-normal  and this motivates the 
choice of our noise model (loss function) in Section \ref{sec:recon}.

\subsection{Two point functions}
\label{sec:perf2pt}

\begin{figure}
\centering
\includegraphics[width=\textwidth]{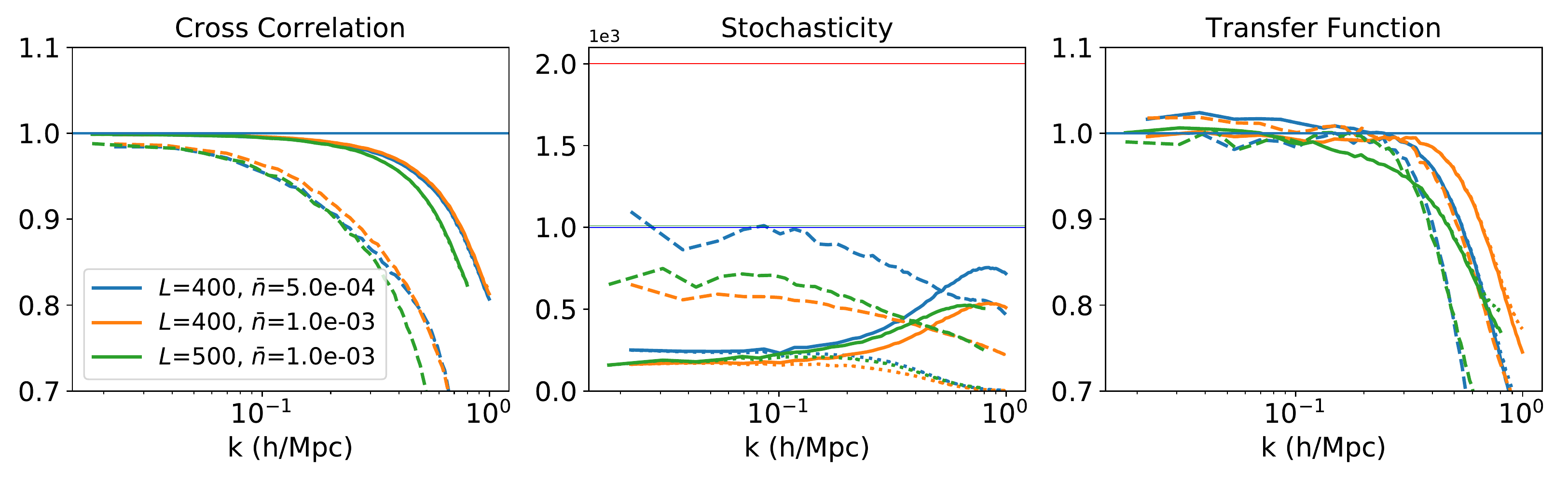}
\caption{Cross correlation coefficient, stochasticity and transfer function for position (dashed) and mass (solids) weighted model field and FOF halos, as well as for the smoothed mass field (dotted). The horizontal lines in the middle panel are the Poisson shot-noise levels for the two number densities.}
\label{fig:2ptmean}
\end{figure}

Finally, we also compare the performance of the model in predicting CIC convolved halo mass and 
halo position fields at the level of 2 point functions in Fourier space. We take reconstructed 
Fourier modes $\delta_N$ and compare them to the true halo modes $\delta_h$. 
To this end, we consider three different metrics - 
\begin{itemize}
\item cross correlation coefficient ($r_c$)
\begin{equation}
r_c = \frac{P_{hN}}{\sqrt{P_{hh}P_{NN}}}
\end{equation}
\item stochasticity ($s$)
\begin{equation}
s = P_{hh} (1 - r_c^2)
\end{equation}
\item and transfer function ($TF$)
\begin{equation}
TF = \sqrt{\frac{P_{NN}}{P_{hh}}}
\end{equation}
\end{itemize}
where $P_{hh}$ is the auto-power spectrum of FOF halo mass (or position) field, $P_{NN}$ is the auto spectra 
for the model (or NNp) and $P_{hN}$ is cross spectrum.
\fig{fig:2ptmean} shows the scale dependence of these statistics for different resolutions and number densities, after taking mean of 5 independent realizations to reduce the noise.
The dashed lines are position weighted FOF halos compared with output of NNp. 
The mass weighted halo fields and the modeled field are compared in solid lines. In dotted lines, we show the comparison between smoothed-mass weighted fields where smoothing is done at $3 \mpch$ with a Gaussian kernel, which is the same field that is used to estimate the error histograms in the previous section.
Poison shot noise is shown as horizontal lines in the middle panel for the two number densities. 

The cross correlation coefficient for position of halos on large scales is $\sim 0.98$ and drops to $\sim 0.9$ at $k = 0.2,\ 0.3 \ihmpc$ for $\bar{n} = 5\times 10^{-4},\ 10^{-3}{\rm (h/Mpc)^3}$ respectively. 
The mass weighted fields are more than $95 \%$ correlated upto $ k \sim 0.7 \ihmpc$  with $r_c$ being impressively close to one upto $k\sim 0.1$ (Mpc/h)$^{-1}$. 
The cross correlation coefficient is much higher for mass weighted fields than the position fields because the former up-weights the heavy halos by orders of magnitude and the model is more accurate in detecting these halos (\fig{fig:missed} and \ref{fig:noise}). 

At the level of stochasticity (against the FOF halo field), the model improves by a factor of 2 over of Poisson shot noise for the position fields. It is a significantly higher reduction as compared to simplistic bias models often used to model halo fields \cite{Modi2017}. This is primarily driven by the fact that NNp is able to correctly identify the neighborhood of halos as discussed in the previous sections. For the mass weighted fields, the stochasticity is much lower than the position field, which is again because its dominated by heavier halos. The stochasticity for the mass weighted field is also flat on large scales and then rises up on small scales. For the smoothed mass field shown in dotted lines however, this increase in stochasticity on smaller scales is suppressed and now stochasticity is white on all the scales that it is non-zero. 

The transfer function of our modeled field with respect to the halo mass field is noisier than either cross-correlation or stochasticity. Further, its close to but not unity on large scales for neither the position nor the mass weighted fields. This likely directly affects the transfer function of our reconstructed field (section \ref{sec:results}) that will need to be calibrated out when reconstructing band powers for linear power spectrum.

\section{Reconstruction}
\label{sec:recon}
In the previous section, we have developed a forward model \bi{f(s)} which does the non-linear mapping to go from the initial linear density modes ($\bi{s}$) to the final halo mass field ($\bi{d}$).
This forward model consists of 5 step \texttt{FastPM} simulation followed by
two neural networks performing non-linear transformations on the evolved density fields to generate our modeled halo mass field.
Given this observation, \bi{d}=$\{d(r_{i(=1...N)})\}$, we are interested in recovering  
the initial density linear modes \bi{s}=$\{s_{j(=1...M)}\}$. In this section we explore this problem,
closely following the formalism developed in \cite{Seljak2017}. An overview of our approach is shown in the left panel of \fig{fig:summary}.
In section \ref{sec:lossfunction}, we use the modeling error discussed in section \ref{sec:perfhist} to develop a loss function that we will optimize over to do the reconstruction. 
Then in section \ref{sec:opt}, we describe the optimization procedure and discuss methods to speed up and improve the convergence. 
In section \ref{sec:results}, we finally present the results for our reconstruction for different number densities. In the next section \ref{sec:impactloss}, we discuss how these results depend on our choice of loss function. 

\subsection{Loss Function}
\label{sec:lossfunction}
We write the observed halo mass field as a combination of signal term (\bi{f(s)}) and a noise term (\bi{n})
as \bi{d = f(s) + n} where the noise, \bi{n}, has contributions from both, noise in the data as well as the  
errors in our model. One can thus write the joint probability of the signal and the noise as a 
product of individual probabilities, under the assumption that they are uncorrelated, as $P(\bi{s, n}) = P(\bi{s}) P(\bi{n})$.

The initial modes are Gaussian and we assume them to be in Fourier space, where their Gaussian prior can be written in a diagonal form. 
\be
P(\bi{s}) = (2\pi)^{-M/2}\rm{det}(\bi{S})^{-1}\rm{exp}\bigg(-\frac{1}{2}\bi{s}\bi{S}^{-1}\bi{s}^\dagger\bigg)
\label{eq:priorpdf}
\ee
with covariance $\bi{S} = \langle \bi{s}\bi{s}^\dagger\rangle$, which is also the power spectrum. We will assume that this power spectrum can be written in terms of parameters $\bi{\Theta}$ which are typically the bandpowers. 
The modes are complex and obey $\bi{s}^{*}(\bi{k}) = \bi{s}(−\bi{k})$, where $\bi{k}$ is the wavevector, but in our labeling of modes we treat real and imaginary component as two independent modes. 

The noise vector $\bi{n}$ is parameterized with the noise covariance matrix $\bi{N} = \langle \bi{n} \bi{n} ^{\dagger} \rangle$.
This noise matrix is made up of 2 contributions - i) measurement noise that is assumed to be known, uncorrelated with
the signal and diagonal (or sparse) in configuration space; ii) modeling noise that may be signal dependent but is 
still diagonal or sparse in configuration space. This consists primarily of the modeling uncertainty caused by our using a substitute
Neural Net model for halo mass and position field in place of actual FOF halo masses and positions.
Contributions to the measurement noise, that we do not include in this work, come from the fact that we do not know the dark matter halo masses accurately. These masses are often calibrated using weak lensing observations and there is 
noise associated with this calibration. In addition, we only observe the halos that host a galaxy and there is an intrinsic scatter associated with the halo mass and galaxy occupancy, as well as relationships between the halo masses and galaxy observables such as stellar mass and luminosity\cite{Rodriguez-Torres2016}.

Our noise model for the modeling error caused by using NN model is based on the error histograms presented in section \ref{sec:perfhist}. 
Our working variable constructed out of our data (and model) for which we estimate error is 
\be
x = \log(M_R + M_0)
\label{eq:var}
\ee 
We estimate error at the level of smoothed halo mass fields for the reasons explained in that section. 
We work with the logarithm of the the mass rather than the mass directly so as to 
decrease the dynamic range over different halo masses
and also make it comparable to the dynamic range of the prior on the initial Gaussian field. 
Furthermore, we add a constant nuisance parameter $M_0$ to the smoothed halo mass field. 
This suppresses the points which have mass assigned less than that value. For instance, 
by assigning a high value to $M_0$, we can suppress the lower mass halos and only compare the fields
at the position of big halos. We can thus tune the value of $M_0$ such that we are affected only 
by the halos above the requisite abundance in our model and exploit this to assist in convergence 
when minimizing the loss function during reconstruction.

In \fig{fig:noise} we have seen that the probability 
distribution of the noise, both with and without the scatter in halo masses, is described well with the displaced log-normal distribution,   
albeit with different values of the parameters (standard deviation and the offset). 
With this noise model, one can then write the probability distribution of noise as 
\be
P(\bi{n}) = (2\pi)^{-N/2}\rm{det}(\bi{N})^{-1} \Pi_i\rm{exp}\bigg[- \bigg(\frac{\mu_N + log(M^{NN}_{R,i} + M_0) - log(M^{FOF}_{R,i} + M_0)}{2\sigma_N}\bigg)^2 \bigg],
\label{eq:probnoise}
\ee
where the product is over all grid cells. This is equivalent to the likelihood of the data. 

Combining this with the prior term (Eq. \ref{eq:priorpdf}) and under the assumption that these are uncorrelated,
one can write the joint probability for the signal and noise as a product of the two. 
This is also the conditional posterior probability of the signal, given the data, up to a constant evidence.
Then, following \cite{Seljak2017}, we can maximize the posterior of the signal to obtain the MAP estimate of initial modes. This can be achieved by instead minimizing the negative log-posterior which is equivalent to minimizing the corresponding $\chi^2$ in the exponent. Thus our loss-function to be minimized is,

\be
\mathcal{L}(\bi{s}) = \frac{1}{2}\bi{s}\bi{S}^{-1}\bi{s}^\dagger + \sum_i\rm{\bigg(\frac{\mu_N + log(M^{NN}_{R,i} + M_0) - log(M^{FOF}_{R,i} + M_0)}{2\sigma_N}\bigg)}^2 \times f_{\rm eff}.
\label{eq:loss}
\ee
where the first term in the prior term and the second is the noise, or the residual term. $\rm f_{eff}$ is the compensating factor which is fraction of effective number of points as discussed below. $\mu_N$ and $\sigma_N$ are the offsets and standard deviations of the displaced log-normal fits to the noise model (\fig{fig:noise}) and depend on $\rm M^{FOF}_R$ at the grid point.

The prior term is diagonal in Fourier space. We expect the noise term to be sparse in data 
(configuration) space and have correlations only on small scales. This is expected because we have smoothed the field to better exploit the information that we are in the correct neighborhood of the halos and hence improve on the gradient directions. Even otherwise, density and hence the mass prediction is correlated in regions around halos, which can lead to correlations in the modeling error.
Using a covariance with non-zero diagonal entires for the noise makes the evaluation of the loss function computationally expensive, which can pose a serious problem for the optimization.

An alternate strategy to handle such covariances is to reduce the  effective number of points in the data set that contribute to the meaningful information, since if ignore these correlations, we overestimate the information content of the residual term. This can be achieved by simply changing the relative weighing of the prior and the residual term in the loss function. 
This simplifies our loss function significantly and leads to the compensating factor $\rm f_{eff}$, which is the fraction of effective number of points on the grid after taking noise correlations into account.
In practice, we have set $\rm f_{eff}$ to 1 and instead used a constant $\sigma_N$ at a level higher than the one estimated by the error histograms (\fig{fig:noise}). This is motivated by the observation that even though the modeling error is smaller for high mass halos (i.e. lower $\sigma_{n}$), the points in this neighborhood are expected to be correlated to larger scales and to a higher degree than the intermediate mass halos (i.e. leading to lower $\rm f_{eff}$). The simplest way to capture this trade-off is to use the constant noise model and we find that it works very well in practice. This is further discussed in the section \ref{sec:impactloss}.

\subsection{Optimization Algorithm with Annealing}
\label{sec:opt}

We do reconstruction of the initial density modes by minimizing the loss function (Eq. \ref{eq:loss}).
This is an optimization problem in a high dimensional space, with both the number of underlying features (initial modes) and the number of data points (grid cells) being in millions or
more. A brute force optimization scheme thus has no real hope of converging. 
At the same time, while second order methods such as Newton's method may perform better than simple gradient descent, Hessian inversion is not feasible. Instead, we balance the trade-offs between the two approaches by using limited memory BFGS (L-BFGS) algorithm which is a quasi-Newton's method, where one approximates the inverse Hessian using low rank matrix approximation constructed from gradients of previous iterations.

L-BFGS algorithm requires gradients of the loss function with respect to all of the initial modes.
This involves evaluating the gradient of the forward model, which in our case consists of a 5 step \texttt{FastPM} evolution followed by transformation of the final density field with Neural networks. 
The latter involves a series of matrix multiplication with the weights of different layers followed by analytic activation functions and hence estimating its derivatives is straightforward. The derivatives of \texttt{\texttt{FastPM}} evolution can be evaluated along the lines of PM derivatives as presented in \cite{Wang2014,Feng2018}. 
We implement these gradients using the automated differentiation package, \texttt{ABOPT}\footnote{\url{https://github.com/bccp/abopt}}, developed with this application in mind.

Since we are aware of physics driving our model as well as its performance, we can use our domain knowledge to assist the convergence by modifying the loss function over the iterations rather than simply brute-forcing the optimization with the vanilla loss function.
With respect to the dynamics of our model, we know that the large scales are linear and posterior 
surface is convex and thus easy to converge, albeit low in number of modes. In terms of the model performance, we know that our model works better for larger mass halos than the smaller ones. Also, optimizing over a discrete field is harder as compared to a smoother, more continuous field. Based on these intuitions, we do the following optimizations. 
\begin{itemize}
\item \emph{Annealing of multigrid smoothing ($N_{sm}$)} \cite{Feng2018}: We smooth the residual term of the loss function on small scales with a Gaussian kernel. 
Thus on these scales, the prior term pulls the small scale power to zero and we force the optimizer to get the large scales correct first. 
Instead of picking a physical scale, we set the smoothing scale in terms of mesh scale by choosing 
$N_{sm} = $ number of grid cells over which to smooth the residual and gradually decrease $N_{sm}$ to 0 (annealing). 
Without multigrid smoothing, we find that the small scale modes overwhelm the optimizer with their sheer number since the number of modes scales as $k^3$. Moreover, since they are more non-linear and more noise contaminated than the large scales which are linear, the optimizer spends most iterations tweaking these small scales without getting the large scales correct.
\item \emph{Annealing of $M_0$}: We start with a high value of $M_0$ and then decrease it over the course of optimization. Since our working variable in the loss function is $\log(M+M_0)$, this suppresses all the points on the grid where FOF halo mass is less than $M_0$. Hence we effectively force the optimizer to first get the large halos correct and then go down in halo mass. 
This helps in converging to the truth faster for two reasons: it  prevents the optimizer from getting overwhelmed with small halos which are more numerous but where our model is slightly worse than it is for more massive halos. Also, the large scale power is driven by the bigger halos due to the mass weighing of the halos and identifying them correctly helps in converging to the correct large scales quicker.

\item \emph{Annealing of discreteness parameter $a$}: The last activation function in the NNp network that gives the position mask is the logistic 
function, Eq. \ref{eq:logistic}, with $a=3$, and this is responsible for the discreteness of the model field. However, discrete fields are harder to optimize over and hence we begin with $a=1$ and then increase its value
to $2$ and $3$ over the iterations. The logistic function saturates gradients when the values are much larger or smaller than the threshold (which is why its no longer favored in the hidden layers of neural networks) and this is more severe with sharper functions ($a=3$). Thus starting with a `broader' ($a=1$) logistic function 
saturates the gradients less, thus assisting in convergence. Physically, it can be understood as an additional smoothing or 
decreasing the discreteness of the modeled field. Since the gradients are non-zero over more grid points, they have a better spatial information in which direction to move initially when we are far from the truth. 
\end{itemize}


\subsection{Results}
\label{sec:results}

In this work, we show the results for reconstruction of initial density field using our forward model in a $400 \mpch$ box over $128^3$ grid for different number densities of halos which were identified in a $512^3$ simulation with 40 time steps. Our optimizer starts with random phases for the initial conditions. The amplitude of the initial conditions is $10\%$ of the fiducial power for high number densities ($\bar{n} = 10^{-3}{\rm (h/Mpc)^3}$ and $5\times 10^{-4}{\rm (h/Mpc)^3}$) and $50\%$ power for the lower number densities ($\bar{n} = 2.5\times 10^{-4}{\rm (h/Mpc)^3}$ and $10^{-4}{\rm (h/Mpc)^3}$).
The correlations in the noise are addressed by changing the relative weighting of the residual and the noise term in the loss function. Instead of using an explicit $\rm f_{eff}$, we do so by using a higher $\sigma_N$ than the mass-dependent noise estimated from the displaced log-normal fits to the error histogram similar to \fig{fig:noise}.
For this subsection, the amplitude (variance) of the noise used is $\sigma_N^2 = 0.25$.
We also keep the non-zero offset ($\mu$) in the means of the error histograms. In the next subsection, we will discuss how these choices in the loss function affect our results.  

To assist the convergence of our optimization, we tweak the loss function over iterations in various ways. We have tried to change the parameters in different order and to multiple values. For the number densities under consideration, the following algorithm provides the optimal balance between run-time and performance-
\begin{itemize}
\item For $M_0 = 10^{12} M_\odot$/h
\begin{itemize}
\item For $N_{sm} \in \{4, 2, 1, 0.5, 0\}$
\begin{itemize}
\item For $a \in \{1, 2, 3\}$
\begin{itemize}
\item do optimization using L-BFGS algorithm until $\mathcal{L}$ (loss function) per degree of freedom stops decreasing by $0.1\%$ over successive iterations
\end{itemize}
\end{itemize}
\end{itemize}
\item For $M_0 = 10^{11} M_\odot$/h \& $a=3$
\begin{itemize}
\item For $N_{sm} \in \{1, 0.5, 0\}$
\begin{itemize}
\item do optimization using L-BFGS algorithm until $\mathcal{L}$ (loss function) per degree of freedom stops decreasing by $0.1\%$ over successive iterations
\end{itemize}
\end{itemize}
\end{itemize}

The values of $M_0$ roughly correspond to halo masses above which lie 
$\sim25\%$ and $\sim50\%$ of the points on the smooth halo mass grid. 
Further, the smallest halo for $\bar{n} = 10^{-3}{\rm (h/Mpc)^3}$ has mass $\sim 5\times10^{12} M_\odot$/h and
using these values of $M_0$ ensures that we have picked the neighborhood of most halos while suppressing most of the empty regions. 
This is useful since once we have converged for $M_0 = 10^{12} M_\odot$/h, we do not need to start again with large smoothing of four grid cells as we have practically converged on the large scales. Also, we do not need to work with a smoother model ($a=1,\ 2$) and can directly use the correct model ($a=3$) for the discrete field. This is because spatially, we have already converged close to the truth.
Decreasing $M_0$ to less than  $10^{11} M_\odot$/h does not improve cross-correlations significantly, likely due to inaccuracies in our model at these small masses and decreasing signal to noise as we go down in mass and scales.

The optimizer takes $\sim 100$ iterations to converge to the best-fit initial field for $M_0 = 10^{12} M_\odot$/h and $\sim 80$ iterations to converge for $M_0 = 10^{11} M_\odot$/h, given our tolerance of $10^{-3}$ i.e. $0.1\%$ decrease in the loss function before declaring convergence. Thus in all, we converge in $\sim 200$ iterations. The number of iterations increases and the performance worsens if we do not assist
the optimizer as outlined above. For instance, when starting with $M_0 = 10^{11} M_\odot$/h instead, the reconstruction takes $\sim 250$ iterations and converges to a worse solution with lower large scale power. 

\begin{figure}
\centering
\includegraphics[width=\linewidth]{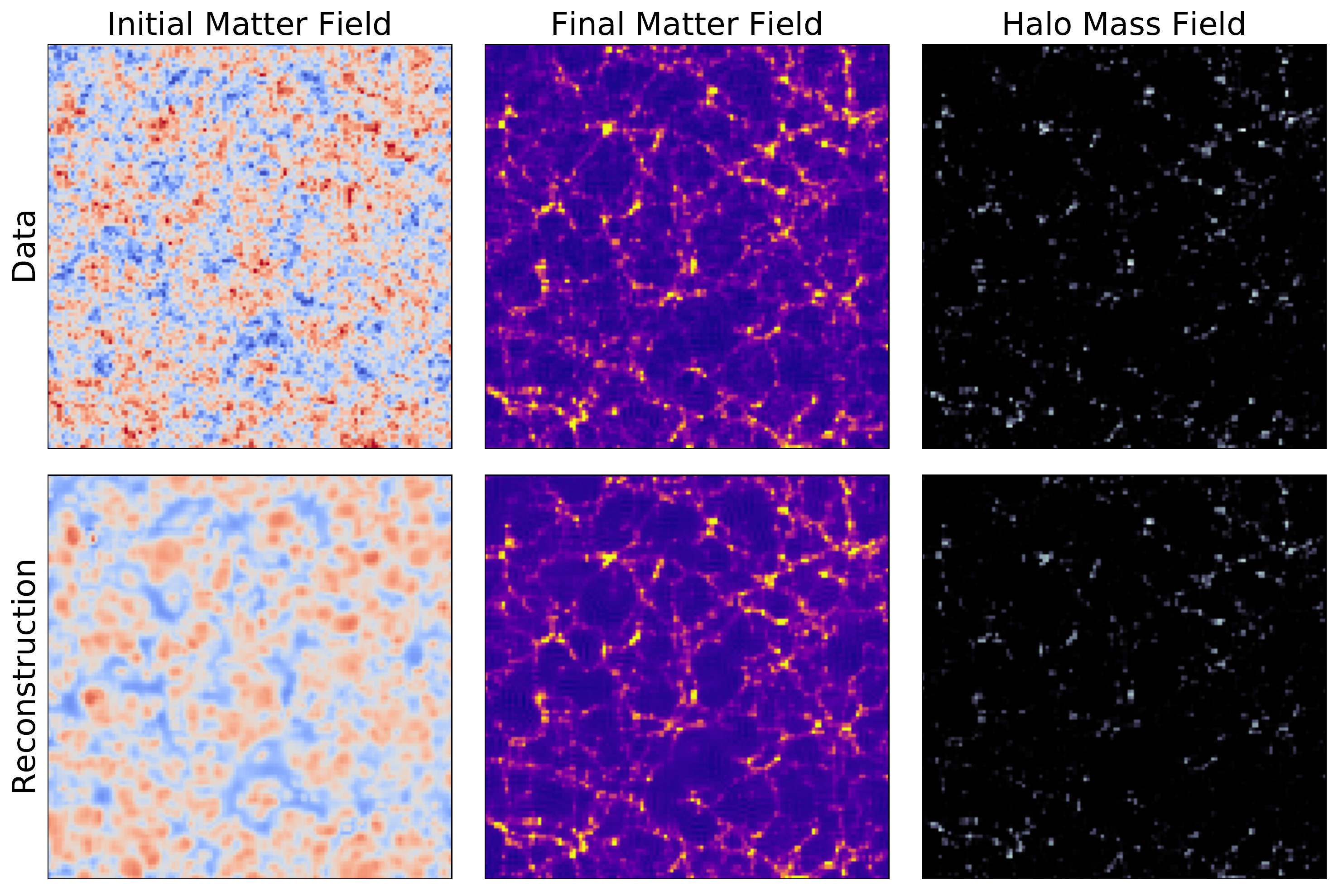}
\caption{Reconstructed fields for $400 \mpch$ box, $128^3$ grid and $\bar{n} = 10^{-3}$ number density at the best-fit (converged) iteration with $M_0 = 10^{11} M_\odot$/h. The color scheme is consistent in every column but different across the columns. The projection is over a slab of thickness $25 \mpch$}
\label{fig:recon2d}
\end{figure}

We first gauge our reconstruction visually in \fig{fig:recon2d} which shows the reconstructed field in the bottom row along with the true field in the top row for the number density of $10^{-3}$. 
The three columns are the initial matter field that we are interested in, final (5 step evolved \texttt{FastPM}) dark matter field, and the halo mass field. For the reconstruction, this halo mass field is the output of the neural networks at convergence while for the data, it is the FOF halo mass field.
At the level of the initial field, we are able to reconstruct the large scale modes quite well while the field is smoothed out on smaller scales.
The small scale suppression is much less severe for the reconstructed Eulerian matter field due to non-linear coupling which transfers power from large scales to small scales. Thus we are able to reconstruct the cosmic web including the filaments, voids and halos remarkably well. The same holds true for the halo mass field. 
The reconstructed halo mass field underestimates some of the heaviest peaks while no such difference is visible in the final matter field. This is due to the non-zero offset $\mu$ in the loss function (see Eq. \ref{eq:loss} and \fig{fig:noise}) which accounts for the difference between the two halo mass fields.

\begin{figure}
\centering
\includegraphics[width=\linewidth]{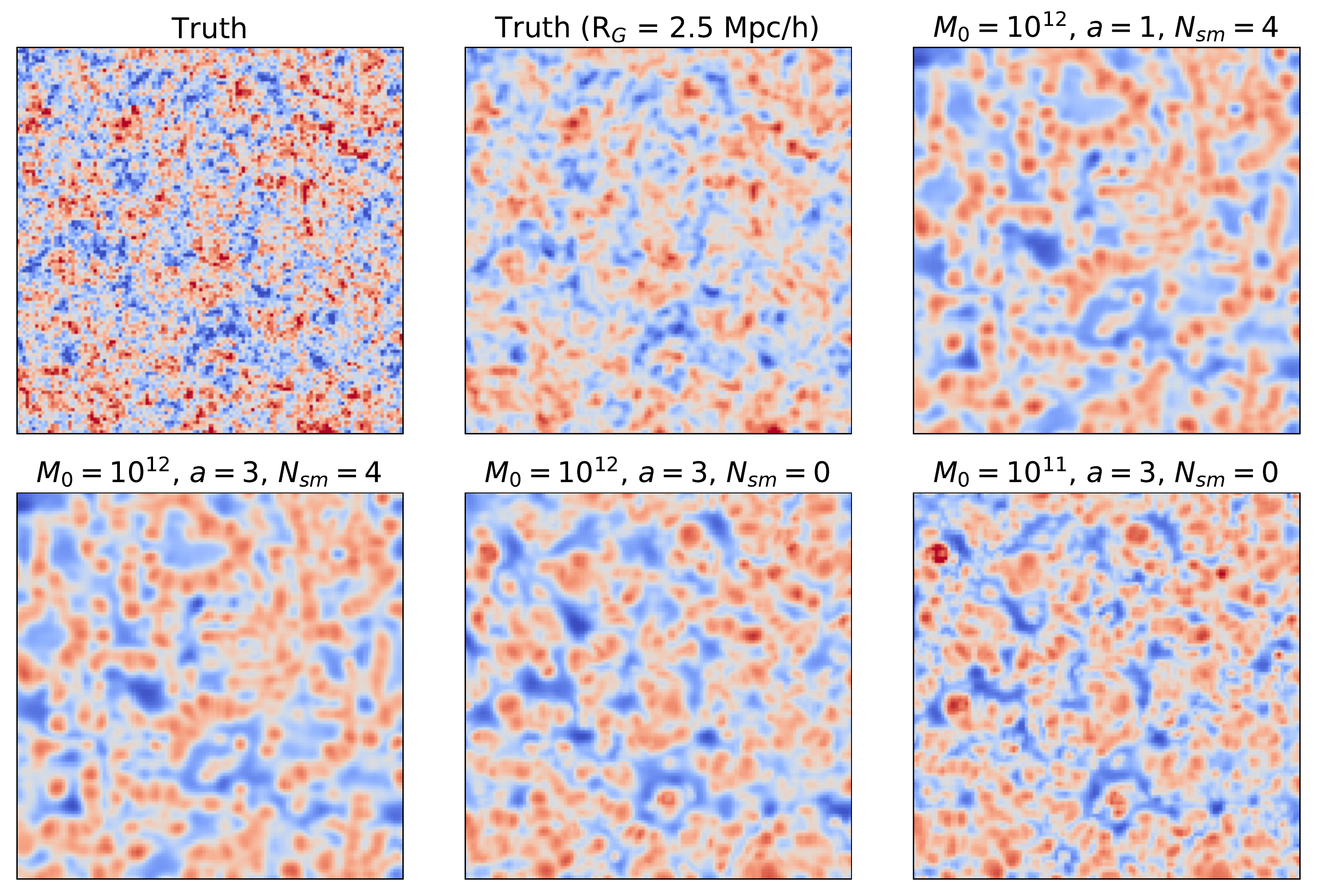}
\caption{Reconstructed initial field at the various stages of the annealing procedure compared with the truth and its smoothed version with Gaussian kernel of $R=2.5\mpch$. The projection is over a slab of thickness $25 \mpch$.}
\label{fig:twelve2d}
\end{figure}

In \fig{fig:twelve2d}, we show the reconstructed initial field at the various stages of the annealing procedure used in the optimization. For comparison, we also show the true initial field and its Gaussian smoothed version with smoothing scale of $R_G =2.5 \ihmpc$. Changing the width of the logistic function ($a$) makes the boundaries of various overdense and underdense regions sharper  by increasing the density contrast between them. Then reducing the smoothing scale $N_{sm}$ matches the reconstructed field to the data on increasingly smaller scales. Since $N_{sm}$ suppresses the residual term, the cross-correlation coefficient between the reconstructed field and the truth is zero on scales smaller than set by it. Lastly, at $M_0 = 10^{12} M_\odot$/h, only $\sim ~25\%$ points effectively contribute to loss function and reducing it to 
$M_0 = 10^{11} M_\odot$/h optimizes over more ($\sim 50\%$) points. This gets contribution from smaller halos and halo environments which start getting reconstructed  and it generates smaller scale features in the reconstructed field. 

\begin{figure}
\centering
\includegraphics[width=\linewidth]{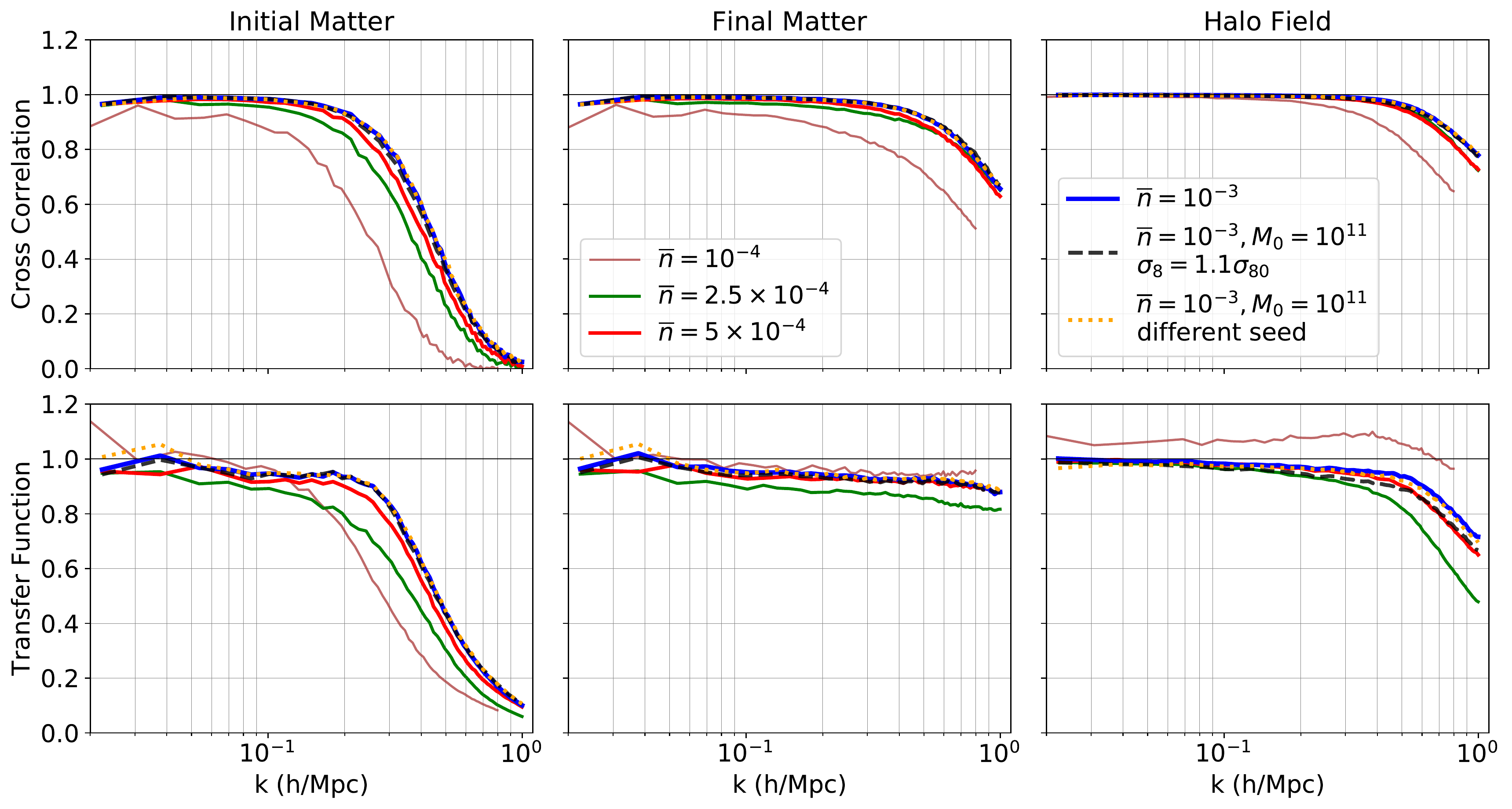}
\caption{Transfer function and cross-correlation between the reconstructed fields and the truth for different number densities, different cosmologies (i.e. different $\sigma_8$) and different realizations (i.e. different phases for the initial condition). Reconstruction for every number density is performed with a different model, trained for that specific number density. However the three $10^{-3} {\rm (h/Mpc)^3}$ number density scenarios (with different seed and cosmology $\sigma_8$) are reconstructed with the same neural network model.}
\label{fig:recon2pt}
\end{figure}

To compare these fields quantitatively, we show the transfer function and the cross-correlation coefficient as function of scale for the reconstructed field in \fig{fig:recon2pt}. We compare the reconstruction for different number densities of halos.
Except for the highest noise case ($\bar{n} = 10^{-4}{\rm (h/Mpc)^3}$), in all cases the cross-correlation for all the three (initial, final and halo) reconstructed fields is very close to unity on large scales. 
With increasing number densities, we are able to push the reconstruction to smaller scales. The cross-correlation for the reconstructed initial field with the truth falls to less than $95\%$ at $k=0.12,\ 0.16,\ 0.18\ \ihmpc$ for $\bar{n} = 2.5 \times 10^{-4},\ 5 \times 10^{-4}\ \&\ 10^{-3}{\rm (h/Mpc)^3}$, respectively. 
For the final matter field, $r_c$ drops to $0.95$ at $k=0.23,\ 0.32,\ 0.42\ \ihmpc$. The cross correlation coefficient of the final matter field is much higher than the initial field on the smaller scales, as also evident from the slices in \fig{fig:recon2d}.

To gauge the cosmology independence of the model and the procedure, we also do reconstruction by changing cosmology: we generate the halo mass field data from the initial conditions with the same phases but different amplitude ($\sigma_8$). Changing $\sigma_8$ does not seem to affect the performance at the level of any of the statistics we look at here. This is despite the fact that the model neural network was trained only for the single, fiducial cosmology.
We also verified that the reconstruction is independent of the realization by doing it for different initial conditions within the same cosmology by changing the seed that changes the phases (and the amplitude within cosmic variance). Both of these are crucial since they will form the basis of our bandpower reconstruction which will enable us to extract the cosmological parameters from this exercise.

\begin{figure}
\centering
\includegraphics[width=\linewidth]{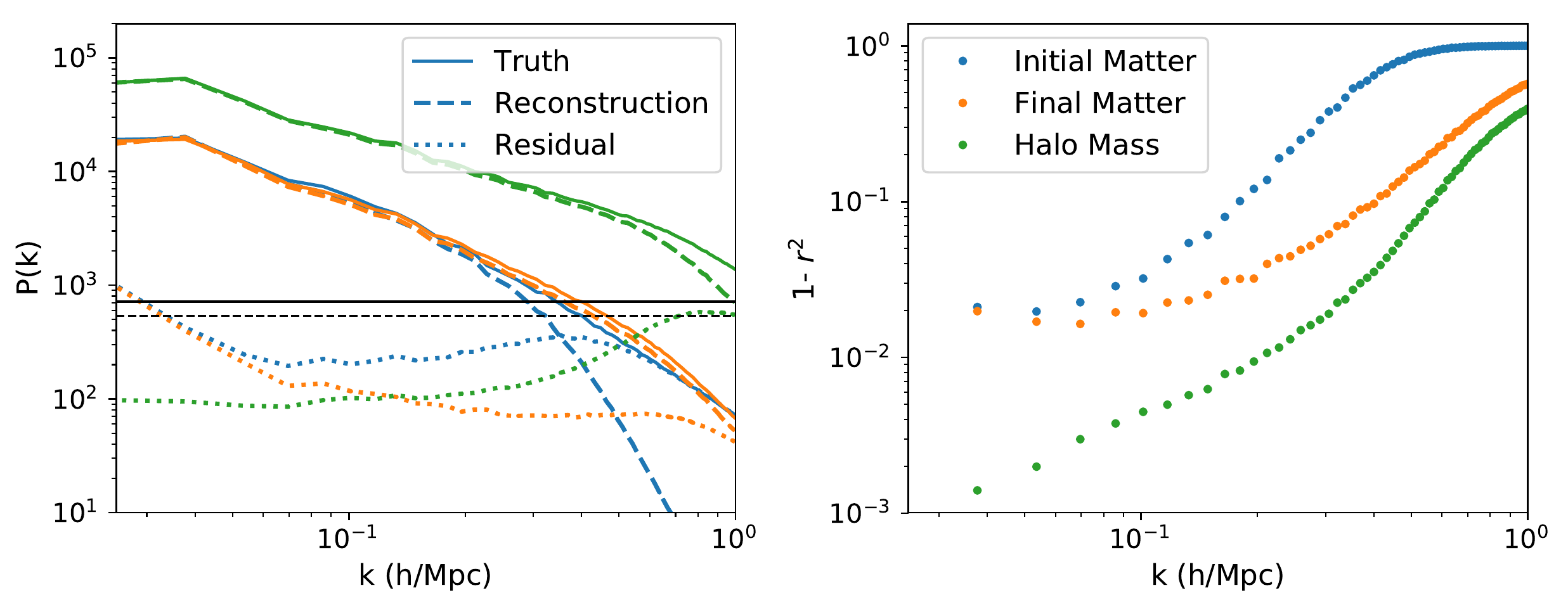}
\caption{Left: Power spectrum of the true (t), reconstructed (r) and the residual (t-r) initial, final and halo mass fields for number density $\bar{n} = 10^{-3}{\rm (h/Mpc)^3}$. The solid horizontal line is noise level in the position field ($\sigma_n/b_n^2$, where $b_n \sim 1.2$ is the bias of the halo position field and $\sigma_n = 1/\bar{n}$) and the dashed line is the noise level in the mass field ($\sigma_m/b_m^2$ where $b_m \sim 1.85$ is the bias of the mass-weighted ($w_i$) field and $\sigma_m = V \times (\sum w^2_i) / (\sum w_i)^2$) with $V$ being the volume. Right: Error power ($1-r^2$, where $r$ is the cross-correlation coefficient of the reconstructed field with the truth) for the three fields.}
\label{fig:powerres}
\end{figure}

One can also estimate the error in the reconstruction by estimating the power of the residual fields, where the residuals are simply the difference of the true and the reconstructed fields. This is shown in the left panel of \fig{fig:powerres} for the reconstruction with number density $\bar{n} = 10^{-3}{\rm (h/Mpc)^3}$. We compare this with the relative noise level in the data (halo mass and halo position fields) by estimating $\sigma/b^2$ where $\sigma$ is the shot noise and $b$ is the bias of the data field. This corresponds to the noise level one would expect if the problem were a linear problem and reconstruction were done with a Weiner filter. We find that in our case, the noise in reconstructed initial field is reduced by a factor of 2 over this noise level.
Another estimate of error in the phases is the quantity $1-r_c^2$, where $r_c$ is the cross-correlation coefficient of the reconstructed field with the true field. This makes it easier to gauge the performance on large scales. We show this in the right panel of \fig{fig:powerres} for $\bar{n} = 10^{-3}{\rm (h/Mpc)^3}$. Note that unlike Figure 4. of \cite{Schmittfull2017}, we do not rescale the reconstructed fields here to have the correct transfer function. 

Unlike the cross-correlation coefficient, the transfer function of the different fields is less well behaved and not unity (\fig{fig:recon2pt}). While one would expect to reconstruct all the power, at least for the halo mass field since that directly enters the loss function, it is not the case. One reason for this is that the field being optimized over in the loss function is a non-linear transformation of the halo mass field which is presented in \fig{fig:recon2pt}. This is to say, we optimize over $x = \log(M_R + M_0)$ as a working variable, not the halo mass field itself, and the Fourier modes do not commute across such transformations. Another reason is the inaccuracy of the NN model itself: we have observed that 
the transfer function was not unity even at the truth, as shown when gauging the performance of the model in \fig{fig:2ptmean}.
In the next subsection, we will show that the transfer function is also more sensitive to the choices we make for the different parameters in the loss function. 

This implies that to reconstruct the linear power spectrum, one cannot simply estimate the power spectrum of the reconstructed initial field. This is well known in the linear case, for instance in Wiener filter reconstruction the reconstructed power is suppressed on noise dominated scales. This calibration and reconstruction of the linear band powers is non-trivial because in addition to marginalizing over all the latent initial modes, it also requires handling various forward model and nuisance parameters that affect the band powers. For example, here we have assumed that we can estimate the halo masses accurately but 
if that is not the case this uncertainty needs to be marginalized over. The noise in the observables and the scatter between the halo mass and galaxy light is another nuisance parameter that needs to 
be marginalized over: we know it is to some extent degenerate with the amplitude of the power spectrum since it changes the bias of the sample \cite{Yoo2012}. Moreover, the surveys live in redshift space and hence one needs to account for the redshift space distortions. 
An approach to reconstruct the band powers is proposed in \cite{Seljak2017} by performing analytic marginalization of the modes around the MAP. Handling this in full detail is beyond the scope of this work and we will pursue it in the future.

\subsection{Impact of the loss function}
\label{sec:impactloss}

It is important to understand how the various assumptions that we have made regarding the loss function affect our reconstruction of the MAP. Specifically, we wish to estimate how sensitive we are to the choices we have made regarding the noise covariance $\sigma_N$, the offset $\mu$ and the free parameter $M_0$ in the loss function.
For this purpose, we will focus on the reconstruction with number density of $\bar{n} = 10^{-3}{\rm (h/Mpc)^3}$ in the $400 \mpch$ box. Our fiducial loss function, for which we presented results in the previous section has the parameters $\mu(M)$ (mass dependent offset), constant noise $\sigma_N^2 = 0.25$ and $M_0 = 10^{11} M_\odot$/h at the considered final iteration.

\begin{figure}
\centering
\includegraphics[width=\linewidth]{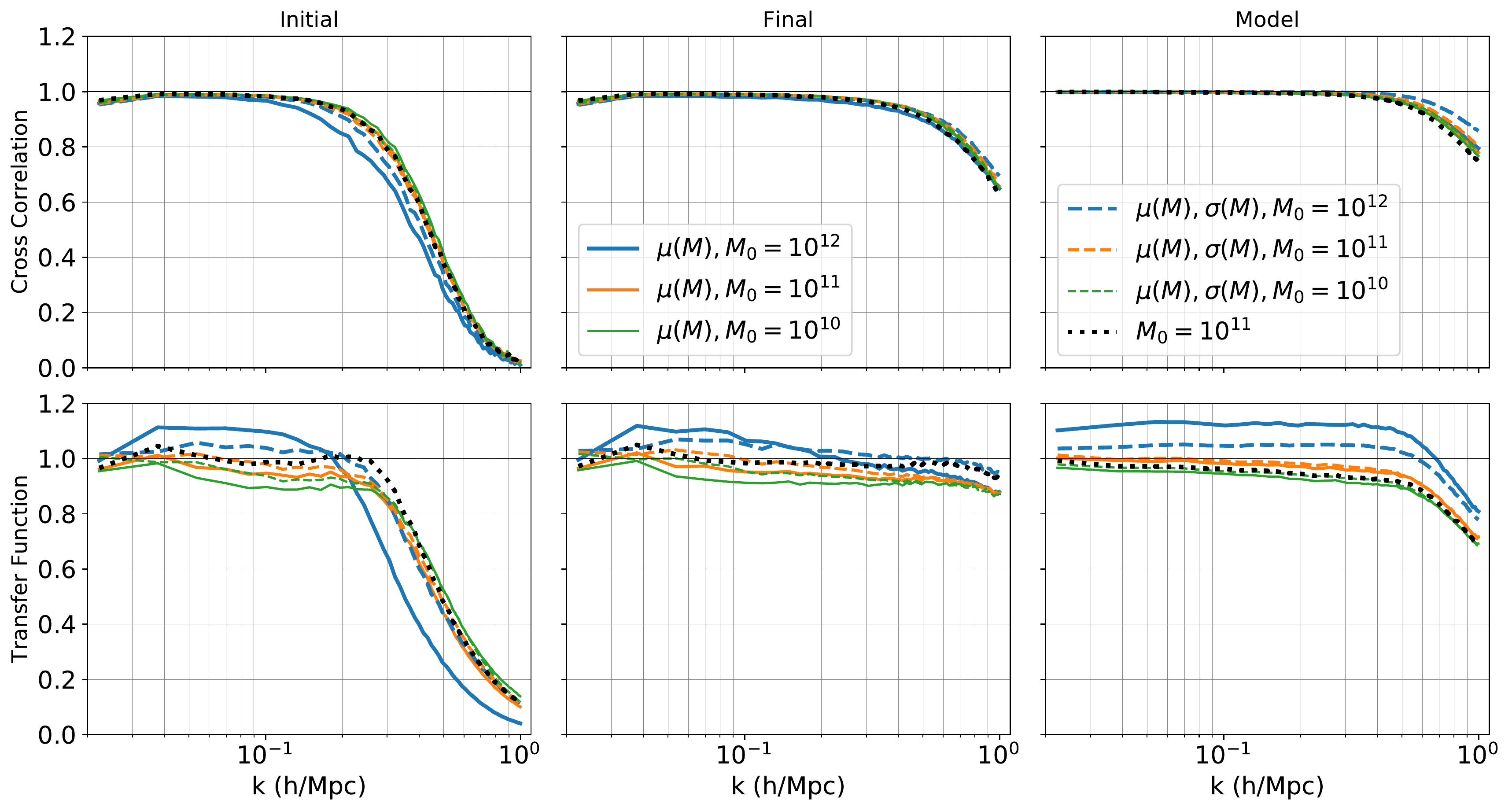}
\caption{Transfer function and cross-correlation between the reconstructed fields and the truth for different parameter choices in the loss function, Eq. \ref{eq:loss}. $\mu(M)$ corresponds to the non-zero offset in the loss-function as estimated from error histograms. $\sigma(M)$ implies use of mass-dependent noise instead of the constant noise as estimated from error histograms. The dotted black line corresponds to the loss function without the offset and constant noise.}
\label{fig:recon2ptloss}
\end{figure}

\fig{fig:recon2ptloss} shows the cross-correlation and transfer function for different choices of the parameters in the loss function. The most direct difference in the performance comes due to the parameter $M_0$. In the previous section, we have advocated beginning with $M_0 = 10^{12} M_\odot$/h and reducing it to $M_0 = 10^{11} M_\odot$/h. For the constant noise model used there, the improvement in cross-correlation of initial fields is significant upon reducing $M_0$  from $10^{12} M_\odot$/h to $10^{11} M_\odot$/h, with the fields correlated to more than $95\%$ upto $k=0.13,\ 0.18 \ihmpc$ respectively. Lowering $M_0$ to $10^{10} M_\odot$/h improves this to $k=0.196 \ihmpc$ but at a greater computational cost.
Interestingly, at the level of the transfer functions, while we lose power on intermediate scales upon decreasing $M_0$ to $10^{10} M_\odot$/h, we gain marginal power on small scales. This is likely due to an increased contribution to the loss function from the small mass halos which drive the small scale power and reduce the overall bias. This increased response might become more important when calibrating band-powers allowing us to push to smaller scales, and we plan to investigate this in the future. 

To handle the correlations in the noise, we have chosen a higher constant noise of $\sigma_N^2 = 0.25$ at all points. Using this noise in comparison to that predicted by the noise histogram reduces the effective number of points contributing to the loss function residual and up-weighs prior. The amount of this down-weighting depends on the value of $M_0$. For $M_0=10^{12}, 10^{11}, 10^{10} M_\odot$/h, using this amplitude of the constant noise, reduces the effective number of points by a factor of $\sim 16, 6, 3$, respectively. As compared to constant noise case, upon using the mass-dependent noise, the cross correlation improves significantly for  $M_0=10^{12} M_\odot$/h but stays the same for $M_0=10^{11} M_\odot$/h, while for $M_0=10^{10} M_\odot$/h, it does not improve over $M_0=10^{11} M_\odot$/h i.e. $95\%$ correlation is at the same $k = 0.18 \ihmpc$ (unlike $k=0.196 \ihmpc$ for constant noise). Moreover, the higher constant noise case converged to the chosen tolerance in $\sim 200$ iterations, while the mass dependent noise took around $\sim 600$ iterations. Thus it seems like we have over-penalized the $M_0=10^{12} M_\odot$/h case with our noise amplitude, but we are adequately handling the $M_0=10^{11}\ \&\ 10^{10} M_\odot$/h loss functions. At this level, we also seem to be somewhat insensitive to the actual value of the error at the position of the halos. One explanation for this is that the noise due to the discreteness of the tracers dominates over the error made in predicting mass. On the same note, changing the amplitude of the noise does not degrade the performance at the level of the cross-correlation as long as we do not over-penalize by reducing the effective number of points by a lot (more than a factor of $\sim 10$). Increasing $\sigma_N$ does slightly decreases the transfer function since it effectively up-weighs the prior, but this still needs to be calibrated out in all cases. 

Lastly, to compare the effect of the offset, we perform a reconstruction without it and using the same constant noise as the fiducial case. This reconstruction is shown in the dotted-black line in \fig{fig:recon2ptloss}. The inclusion or exclusion of the offset does not seem to affect cross-correlation. It might seem like removing offset gives transfer function of unity on large scales, but this case here is a coincidence. Repeating the exercise for other number densities confirms that not including the offset generally gives higher transfer function than when including it, but not necessarily unity. However we prefer to include non-zero offset since preliminary results show that it will become more important when we add signal noise to the halo masses of the data. 

Overall, we find that our reconstruction is most sensitive to $M_0$ with most of the information being reconstructed by $M_0=10^{11} M_\odot$/h for the number densities studied in this paper, and marginal improvements on small scales after that. Cross-correlation seems to be quite robust against the choice of various other parameters of the loss function. The transfer function is more sensitive to this choice since the parameters mainly serve to change the balance between the residual and the prior, or the amplitude of the matter density fields that is regressed over to give the halo mass (via offset). However note that these things generally shift the transfer function up or down as a whole, with almost no change in the scale dependence. As discussed, since none of these cases result in a response that is unity, we need to calibrate band-powers anyways. Thus we expect that this sensitivity of the transfer function on the parameters of the loss function will automatically be handled during the band-power calibration.

\section{BAO Information}
\label{sec:information}

In this section, we compare our reconstruction against standard reconstruction \cite{Eisenstein2007}. However before doing so, it is important to note that the objective of standard reconstruction is to maximize signal to noise for baryon acoustic oscillations (BAO) and it does not focus on the reconstructing the initial density field map or broadband power.
On the other hand, our goal is to reconstruct the linear density map and via that, reconstruct the linear power spectrum to extract cosmological parameters. 
Nevertheless, at this stage we choose to compare against the standard reconstruction since it is the most widely used method of reconstruction and we leave the band power reconstruction for future work.
Given the differences in the goal of standard and our reconstruction, we will compare the two reconstructions using metrics of the former. We are interested in estimating and comparing the linear BAO information reconstructed in the fields and then using to estimate the fractional error in identifying the location of the baryonic peak using both the methods. We use Fisher formalism for this estimation since our box size is insufficient to recover BAO signal due to the cosmic variance.

\begin{figure}
\centering
\includegraphics[width=\textwidth]{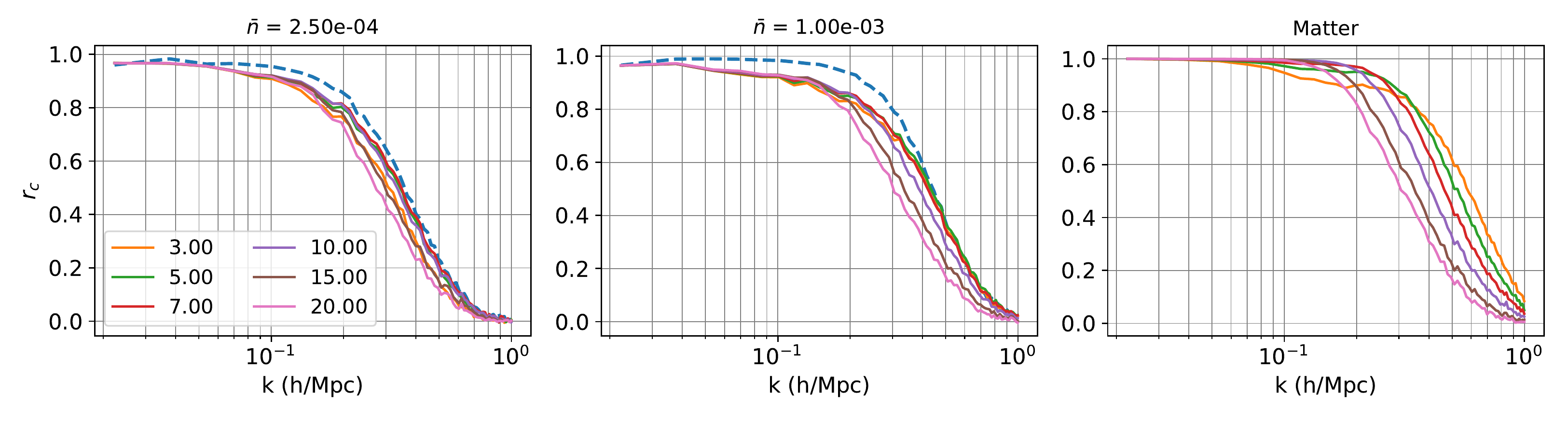}
\caption{Performance of standard reconstruction when done with matter and halos of different number densities for different smoothing scales}
\label{fig:bao_smoothing}
\end{figure}

\begin{figure}
\centering
\includegraphics[width=\textwidth]{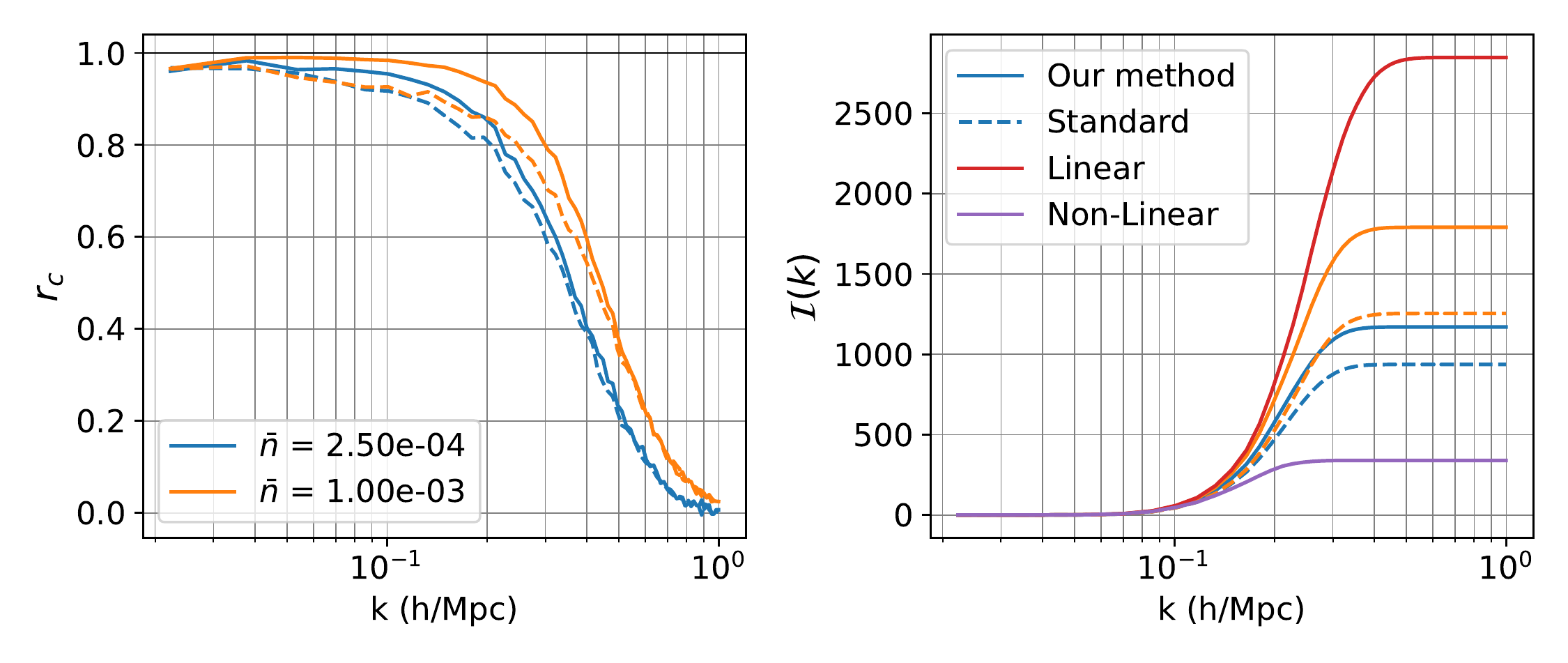}
\caption{Left: Cross correlation coefficient of reconstructed fields from standard reconstruction (dashed) and our reconstruction (solids) with the true initial field when done from data with two different number densities. Right: Cumulative Fisher information to estimate the BAO peak as a function of scale for each case as calculated with Eq. \ref{eq:Fishers0}. For comparison, we also show the same Fisher information for linear matter field (theoretical maximum) and the non-linear matter field.}
\label{fig:snr_lin}
\end{figure}

To recall, we work in a box of L $= 400 \mpch$ on a $128^3$ grid and use the halo catalogs with number density $2.5 \times 10^{-4}{\rm (h/Mpc)^3}$ and $ 10^{-3} {\rm (h/Mpc)^3}$  as our data.
Standard reconstruction involves estimating the displacement field using the halo overdensity field. To suppress small-scale non-linearities where linear theory is not valid, we smooth the halo overdensity field with a Gaussian smoothing kernel $S(k) = e^{-k^2\Sigma_{sm}^2 /4}$. The performance of standard reconstruction depends on the choice of the smoothing scale.
Larger smoothing scales are more optimal for lower number densities which are noise dominated on smaller scales (\fig{fig:bao_smoothing}), but smaller smoothing scales reconstruct more information. We choose $\Sigma_{sm} = 10 \mpch$ for $\bar{n} = 2. 5\times 10^{-4}{\rm (h/Mpc)^3}$ and $\Sigma_{sm} = 7 \mpch$ for $\bar{n} = 1\times 10^{-3}{\rm (h/Mpc)^3}$.
We show the cross-correlation coefficient for both, our NN reconstruction and the standard reconstruction with the true linear density field in the left panel of \fig{fig:snr_lin}. Our reconstruction outperforms the standard reconstruction on all scales, with the two fields being $90\%$ correlated respectively up to $k = 0.25 \ihmpc$ and $0.12 \ihmpc$, for number density $\bar{n} = 1\times 10^{-3}{\rm (h/Mpc)^3}$ and $k = 0.16 \ihmpc$ and $0.11 \ihmpc$ for $\bar{n} = 2.5\times 10^{-4}{\rm (h/Mpc)^3}$. 

In BAO community, the linear signal to noise in the reconstructed field is measured in terms of propagator \cite{Seo2010,Seo2016,Yu2017}, defined as 
\begin{equation}
G(k) = \frac{\langle \delta_f \delta_{\rm{lin}}\rangle}{b\ \langle \delta_{\rm{lin}} \delta_{\rm{lin}}\rangle}
\label{eq:propagator}
\end{equation}
where $\delta_f$ is any field whose propagator we wish to evaluate (so it is the reconstructed fields in our case); $b$ is the linear bias of the tracer field (we set $b=1$ for the NN reconstructed field since the bias is effectively modeled by our model of halo field i.e. the neural networks) and $\delta_{\rm{lin}}$ is the initial (linear) field. 
The propagator effectively estimates the projection of the field ($\delta_f$) on the linear density field $\delta_{\rm{lin}}$ \cite{Seo2010}. Then, the linear signal power in the reconstructed field is simply $b^2 G^2(k)P_{\rm{lin}}$. The residual power not captured in the linear signal constitutes the noise term, otherwise referred to as the mode coupling term. Thus this mode-coupling term $P_{\rm{MC}}$ is such that

\begin{equation}
P_{\delta_f} = b^2 G^2(k) P_{\rm{lin}} + P_{\rm{MC}}
\label{eq:mode_coupling}
\end{equation}

Therefore, assuming a log-normal error, the signal to noise ratio for the linear information in the reconstructed field relative to the linear density field is 

\begin{equation}
\frac{S}{N} = \frac{b^2 G^2(k) P_{\rm{lin}}}{b^2 G^2(k) P_{\rm{lin}} + P_{\rm{MC}}} = r_c^2
\label{eq:snr_k}
\end{equation}
where $r_c$ is the cross correlation coefficient of the field with the linear field. 

Given the propagator, we can also estimate the error in locating the BAO peak in the reconstructed signal by doing a simple Fisher analysis. For this, we follow the procedure presented in \cite{Seo2007}. Briefly, assuming Gaussian likelihood for band powers of power spectrum, the Fisher matrix is approximately \cite{Tegmark1997},

\begin{equation}
F_{ij} =  V_{\rm{eff}}\int_{k_{min}}^{k_{max}} \frac{\partial {\rm ln} P(k)}{\partial p_i} \frac{\partial {\rm ln} P(k)}{\partial p_j}\frac{4 \pi k^2}{2 (2\pi)^3} dk
\label{eq:Fisher}
\end{equation}

where we have integrated over the angles assuming isotropy, $p_i, p_j$ are the parameters of interest and $V_{eff}$ is the effective survey volume. The parameter of interest is the location of the centroid of the BAO peak and it corresponds to the sound horizon `$s_0$' at the drag epoch. This is only sensitive to the baryonic component of the power spectrum. BAO peak is damped due to the Silk damping and non-linear evolution and the information lost in the latter damping is modeled well with a Gaussian damping model (\cite{Noh2009}) in theory and the propagator defined above in practice. Reconstruction aims at undoing this non-linear damping and its success is measured by using the propagator estimated for the reconstructed field. Modeling the sensitivity of the power spectrum to sound horizon in this way leads to the Fisher error given by (for full derivation, see \cite{Seo2007,Seo2016}) 

\begin{equation}
F_{{\rm ln} s_0} =  V_{\rm{survey}} A_0^2 \int_{k_{min}}^{k_{max}} \frac{e^{-2(k \Sigma_s)^{1.4}} G^4(k)}{(P(k)/P_{0.2})^2} k^2 dk = \bigg(\frac{s_0}{\sigma_{s_0}}\bigg)^2
\label{eq:Fishers0}
\end{equation}

where $A_0$ = 0.4529 for WMAP1 cosmology, $\Sigma_s \sim 7.76 \ihmpc$ is the Silk damping scale and $P_{0.2}$ is the linear power spectrum at $k=0.2 \ihmpc$. While WMAP1 is not the cosmology of our simulations, our goal here is simply to broadly compare our reconstruction with the standard reconstruction and not to do an accurate Fisher forecast, hence any reasonable value of these parameters should suffice.

We show this Fisher information as the function of scale in right panel of \fig{fig:snr_lin} for the standard reconstruction, and our method for two number densities. For comparison, we also show the Fisher information in the linear and the non-linear matter field which roughly correspond to the maximum and minimum (for matter) information bounds. As expected, at increasing number densities, the standard reconstruction becomes increasingly sub-optimal and we gain $33\%$ and $45\%$ more information at $\bar{n} = 2.5 \times 10^{-4}{\rm (h/Mpc)^3}$ and $\bar{n} = 10^{-3}{\rm (h/Mpc)^3}$ number densities. In effect, we gain as much by using an optimal reconstruction as we do by increasing the number density from  $\bar{n} = 2.5 \times 10^{-4}{\rm (h/Mpc)^3}$ by a factor of 4. 
We find that for this volume the fractional error in locating the peak of the BAO from non-linear field (i.e. estimated using the propagator for the final matter field) is $\sigma_{s_0}/s_0\sim 5.4\%$. By undoing the non-linear evolution, this is reduced to $\sigma_{s_0}/s_0 \sim 3.4\%$ (standard) and $\sigma_{s_0}/s_0 \sim 2.9\%$ (our method) for $\bar{n} = 2.5\times 10^{-4}{\rm (h/Mpc)^3}$ and $\sigma_{s_0}/s_0 \sim 2.9\%$ (standard) and $\sigma_{s_0}/s_0 \sim 2.3\%$ (our method) for $\bar{n} = 10^{-3}{\rm (h/Mpc)^3}$, respectively. Thus our reconstruction method shrinks the error on the location of the peak of BAO by about $\sim 15-20\%$ over the standard reconstruction. However for completeness, it should be noted that this only a Fisher prediction for halos in real space, at $z=0$, and we have assumed that we have the information of halo masses accurately. Thus while the improvement is quite encouraging, we plan to address a more realistic 
case of galaxies in redshift space in the future.

\section{Conclusions and Discussion}
\label{sec:discussion}

In this paper we address the question of how to model a discrete set of measured galaxy positions and their masses and represent them in a form that is suitable to be used for reconstruction of the initial density field using gradient based optimization methods.
We simplify the problem and look at the dark matter halo centers as a proxy for the galaxy position and dark matter halo mass as a proxy for the galaxy stellar mass or luminosity. We also work in real space instead of redshift space where the actual survey data lives. 
Discrete objects are problematic for gradient based methods and halos are typically defined in an N-body simulation using non-differentiable methods such as Friends-of-Friends. 
Since our goal is to find a suitable differentiable forward model, we propose a neural network solution that represents halo properties (position and 
mass) with a set of NN coefficients and activation functions that take in dark matter properties as the input and output quantities such as the probability of finding a halo at a given position and the halo mass (\fig{fig:summary}).
Since NN are explicitly differentiable, once they are trained they can be used as part of the forward 
model of LSS that starts from the linear density modes, evolves them using nonlinear N-body simulation (in our case \texttt{FastPM} \cite{Feng2016}) and 
ends with the predicted galaxy positions and masses. 

We gauge the performance of our forward model of two neural networks in modeling the halo mass and position field on various metrics (section \ref{sec:perf}). We are able to reduce the stochasticity in halo positions over the Poisson shot noise by at least a factor of two and find good cross correlation coefficient up to small scales (\fig{fig:2ptmean}). This suggests that the forward model can be used to generate mock galaxy catalogs, but in this paper we are more interested in using it to reconstruct the initial density field given the data and leave investigating other applications for the future.
To do the reconstruction, one must also define a loss function which takes into account the error probability distribution of the data given the model. For this, we adopt a simple displaced log-normal model, which seems to reproduce the error PDF well (Figure \ref{fig:noise}). Since our model localizes the neighborhood of halos quite well, estimating this error on smoothed mass fields results in white noise spectrum for error. Even though we expect our noise to be correlated on small scales, especially around halo positions, we assume diagonal noise in configuration space and address these correlations by changing the effective number of points contributing to the likelihood, thus altering the relative weighting of the residual and the prior term in the noise model.  

Using this data noise model (Eq. \ref{eq:probnoise}) and Gaussian prior for the modes we are able to do reconstruction of initial density field starting from halo mass field by optimizing the loss function of Eq. \ref{eq:loss} using L-BFGS algorithm. Due to the discrete nature of the data 
we develop several annealing 
methods where we typically gradually change the loss function with number of iterations to increase the resolution from coarse to fine, which help 
nudge the solution towards its final result. These are
motivated by the domain knowledge of the distribution of our data, performance of our model as well as non-linearities generated by the cosmological evolution. This results in a significant speed up and improves the convergence of our optimizer. Specifically, we force the optimizer to fit for the large scales and the massive halos before gradually reducing the scale to fit for smaller scales and smaller halos. We also change the discreteness of the halo field to get better gradients when far from the truth, gradually making it sharper and sharper. 
With about $\sim 200$ iterations we reconstruct the initial density field that is more than $95\%$ correlated to the true field up to $k=0.12,\ 0.16,\ 0.18\ \ihmpc$ for $\bar{n} = 2.5 \times 10^{-4},\ 5 \times 10^{-4}\ \&\ 10^{-3}{\rm (h/Mpc)^3}$, respectively. 
(\fig{fig:recon2pt}). Due to the non-linear mode coupling introduced by gravitational evolution, the reconstructed final matter field has better cross-correlation on smaller scales with $r_c$ dropping to $0.95$ at $k=0.23,\ 0.32,\ 0.42\ \ihmpc$ for the three number densities, respectively. As a result of the reconstruction of small scale power, we are able to identify cosmic web along with all the structures like nodes, voids and filaments in this field (\fig{fig:recon2d}).

Using simulations with different initial conditions (phases) and different cosmology ($\sigma_8$), we also verify that our model and procedure is independent of realization and cosmology. We do a case-study to establish how our reconstruction depends on the choice of each parameter that enters the loss function (\fig{fig:recon2ptloss}). We reconstruct most of the scales by $M_0 = 10^{11} M_\odot$/h and decreasing $M_0$ further changes small scales only slightly.  The cross-correlation coefficient of the reconstructed initial field is fairly robust against the values of the offset ($\mu$) and the noise ($\sigma_N$). Moreover, for $M_0 = 10^{11} M_\odot$/h and lower, using a constant, but higher value of noise, handles the noise covariances adequately. The transfer function seems to be more sensitive to the choices of these parameters, but only in amplitude and not in the scale dependence. Thus we expect to be able to marginalize over these dependencies when calibrating the transfer function to reconstruct the band powers and extracting cosmological parameters, but we leave this for the future. Using our fiducial loss function for reconstruction, we also show that we are able to improve upon standard reconstructions in terms of linear information and localization of BAO peak (\fig{fig:snr_lin}). We expect our method 
should be close to optimal for BAO reconstruction. 

The ultimate goal of our exercise is to construct a tractable summary statistic from which all of the cosmological information can be extracted. For linear modes such statistic is the band powers 
of the power spectrum, and by linearizing the mode evolution we expect we can achieve the same by 
using the reconstructed initial density modes to reconstruct the linear band powers. This is not trivial given that the transfer function for reconstructed field is not unity on all scales. We leave this calibration for the upcoming work. In addition to marginalizing over the latent initial modes, reconstructing band powers also requires that the nuisance parameters of our forward model and data are marginalized over. For instance, noise in the halo mass calibration and scatter in the halo mass-luminosity
relation are both known to be correlated with the amplitude of the power spectrum $\sigma_8$ \cite{Yoo2012}. Thus, a full analysis to obtain cosmological parameters 
must also include a proper marginalization over the nuisance parameters of galaxy formation models. 

This paper shows that one can build a realistic and differentiable halo field from the dark matter density alone, and that
loss functions can be defined that give realistic penalty loss when applying such models to the data. 
To make our modeling more realistic and applicable to the real LSS survey data we must do several additional steps. First, we must 
include additional nuisance parameters such as the satellite distribution inside the halos etc., similar to the 
traditional HOD models \cite{Berlind2002}. Furthermore, we must also include the redshift space distortions by adding velocity to the 
halo position and map it to the redshift space. This will require deriving the gradient of the final 
velocity with respect to the initial modes. 
Redshift space distortions will not only modify the 
forward model 
but will also change the loss function: in particular, the velocity dispersion of the satellites will need to be modeled as an additional 
noise term in the radial direction, since it is unlikely that we can forward model it. Finally, we must also include survey mask effects and 
various galaxy selection effects for a complete forward model. We plan to pursue these issues in the future. The hope is that when a 
complete forward model is available, with a sufficient number of nuisance parameters to describe all the complications not included in 
the noise, then the inverse problem will be completely determined and an optimal analysis of cosmological parameters from the redshift space 
distortions data will be possible. 

\acknowledgments

CM would like to thank Vanessa B\"{o}hm, Yin Li, Patrick McDonald, Ambrish Rawat and Michael Wilson for useful discussions. 
We would also like to thank Vanessa B\"{o}hm, Yin Li, Matias Zaldarriaga and Hong-Ming Zhu for comments on the manuscript. This research used resources of the National Energy Research Scientific Computing Center, a DOE Office of Science User Facility supported by the Office of Science of the U.S. Department of Energy under Contract No. DE-AC02-05CH11231. We acknowledge support of NASA grant NNX15AL17G.

\bibliographystyle{revtex}
\bibliography{sigmoid}

\begin{thebibliography}{10}
\providecommand*{\bibinfo}[2]{#2}
\providecommand*{\eprint}[1]{#1}
\providecommand*{\url}[1]{#1}
\bibitem{Eisenstein2007}
\bibinfo{author}{D.~J. {Eisenstein}}, \bibinfo{author}{H.-J. {Seo}},
  \bibinfo{author}{E.~{Sirko}}, and \bibinfo{author}{D.~N. {Spergel}},
  \bibinfo{journal}{\apj} \bibinfo{volume}{\textbf{664}}, \bibinfo{pages}{675}
  (\bibinfo{date}{Aug. 2007}), \eprint{astro-ph/0604362}.
\bibitem{Zhu2017}
\bibinfo{author}{H.-M. {Zhu}}, \bibinfo{author}{Y.~{Yu}},
  \bibinfo{author}{U.-L. {Pen}}, \bibinfo{author}{X.~{Chen}}, and
  \bibinfo{author}{H.-R. {Yu}}, \bibinfo{journal}{\prd}
  \bibinfo{volume}{\textbf{96}}(12), \bibinfo{pages}{123502},
  \bibinfo{eid}{123502} (\bibinfo{date}{Dec. 2017}), \eprint{1611.09638}.
\bibitem{Schmittfull2017}
\bibinfo{author}{M.~{Schmittfull}}, \bibinfo{author}{T.~{Baldauf}}, and
  \bibinfo{author}{M.~{Zaldarriaga}}, \bibinfo{journal}{\prd}
  \bibinfo{volume}{\textbf{96}}(2), \bibinfo{pages}{023505},
  \bibinfo{eid}{023505} (\bibinfo{date}{Jul. 2017}), \eprint{1704.06634}.
\bibitem{Jasche2013}
\bibinfo{author}{J.~{Jasche}} and \bibinfo{author}{B.~D. {Wandelt}},
  \bibinfo{journal}{\mnras} \bibinfo{volume}{\textbf{432}},
  \bibinfo{pages}{894} (\bibinfo{date}{Jun. 2013}), \eprint{1203.3639}.
\bibitem{Kitaura2013}
\bibinfo{author}{F.-S. {Kitaura}}, \bibinfo{journal}{\mnras}
  \bibinfo{volume}{\textbf{429}}, \bibinfo{pages}{L84} (\bibinfo{date}{Feb.
  2013}), \eprint{1203.4184}.
\bibitem{Wang2014}
\bibinfo{author}{H.~{Wang}}, \bibinfo{author}{H.~J. {Mo}},
  \bibinfo{author}{X.~{Yang}}, \bibinfo{author}{Y.~P. {Jing}}, and
  \bibinfo{author}{W.~P. {Lin}}, \bibinfo{journal}{\apj}
  \bibinfo{volume}{\textbf{794}}, \bibinfo{pages}{94}, \bibinfo{eid}{94}
  (\bibinfo{date}{Oct. 2014}), \eprint{1407.3451}.
\bibitem{Seljak2017}
\bibinfo{author}{U.~{Seljak}}, \bibinfo{author}{G.~{Aslanyan}},
  \bibinfo{author}{Y.~{Feng}}, and \bibinfo{author}{C.~{Modi}},
  \bibinfo{journal}{\jcap} \bibinfo{volume}{\textbf{12}}, \bibinfo{pages}{009},
  \bibinfo{eid}{009} (\bibinfo{date}{Dec. 2017}), \eprint{1706.06645}.
\bibitem{Berlind2002}
\bibinfo{author}{A.~A. Berlind} and \bibinfo{author}{D.~H. Weinberg},
  \bibinfo{journal}{The Astrophysical Journal}
  \bibinfo{volume}{\textbf{575}}(2), \bibinfo{pages}{587}
  (\bibinfo{date}{2002}), \url{http://stacks.iop.org/0004-637X/575/i=2/a=587}.
\bibitem{Wang2009}
\bibinfo{author}{H.~{Wang}}, \bibinfo{author}{H.~J. {Mo}},
  \bibinfo{author}{Y.~P. {Jing}}, \bibinfo{author}{Y.~{Guo}},
  \bibinfo{author}{F.~C. {van den Bosch}}, and \bibinfo{author}{X.~{Yang}},
  \bibinfo{journal}{\mnras} \bibinfo{volume}{\textbf{394}},
  \bibinfo{pages}{398} (\bibinfo{date}{Mar. 2009}), \eprint{0803.1213}.
\bibitem{Wang2013}
\bibinfo{author}{H.~{Wang}}, \bibinfo{author}{H.~J. {Mo}},
  \bibinfo{author}{X.~{Yang}}, and \bibinfo{author}{F.~C. {van den Bosch}},
  \bibinfo{journal}{\apj} \bibinfo{volume}{\textbf{772}}, \bibinfo{pages}{63},
  \bibinfo{eid}{63} (\bibinfo{date}{Jul. 2013}), \eprint{1301.1348}.
\bibitem{Yu2017}
\bibinfo{author}{Y.~{Yu}}, \bibinfo{author}{H.-M. {Zhu}}, and
  \bibinfo{author}{U.-L. {Pen}}, \bibinfo{journal}{\apj}
  \bibinfo{volume}{\textbf{847}}, \bibinfo{pages}{110}, \bibinfo{eid}{110}
  (\bibinfo{date}{Oct. 2017}), \eprint{1703.08301}.
\bibitem{Cautun2011}
\bibinfo{author}{M.~C. {Cautun}} and \bibinfo{author}{R.~{van de Weygaert}},
  \bibinfo{title}{\emph{{The DTFE public software: The Delaunay Tessellation
  Field Estimator code}}}, Astrophysics Source Code Library (\bibinfo{date}{May
  2011}), \eprint{1105.0370}.
\bibitem{Feng2018}
\bibinfo{author}{Y.~Feng}, \bibinfo{author}{U.~Seljak}, and
  \bibinfo{author}{M.~Zaldarriaga}, \bibinfo{journal}{\textit{In preparation}}
  .
\bibitem{Mandelbaum2006}
\bibinfo{author}{R.~{Mandelbaum}}, \bibinfo{author}{U.~{Seljak}},
  \bibinfo{author}{G.~{Kauffmann}}, \bibinfo{author}{C.~M. {Hirata}}, and
  \bibinfo{author}{J.~{Brinkmann}}, \bibinfo{journal}{\mnras}
  \bibinfo{volume}{\textbf{368}}, \bibinfo{pages}{715} (\bibinfo{date}{May
  2006}), \eprint{astro-ph/0511164}.
\bibitem{Auld2007}
\bibinfo{author}{T.~{Auld}}, \bibinfo{author}{M.~{Bridges}},
  \bibinfo{author}{M.~P. {Hobson}}, and \bibinfo{author}{S.~F. {Gull}},
  \bibinfo{journal}{\mnras} \bibinfo{volume}{\textbf{376}},
  \bibinfo{pages}{L11} (\bibinfo{date}{Mar. 2007}), \eprint{astro-ph/0608174}.
\bibitem{Schmelzle2017}
\bibinfo{author}{J.~{Schmelzle}}, \bibinfo{author}{A.~{Lucchi}},
  \bibinfo{author}{T.~{Kacprzak}}, \bibinfo{author}{A.~{Amara}},
  \bibinfo{author}{R.~{Sgier}}, \bibinfo{author}{A.~{R{\'e}fr{\'e}gier}}, and
  \bibinfo{author}{T.~{Hofmann}}, \bibinfo{journal}{ArXiv e-prints}
  (\bibinfo{date}{Jul. 2017}), \eprint{1707.05167}.
\bibitem{Mustafa2017}
\bibinfo{author}{M.~{Mustafa}}, \bibinfo{author}{D.~{Bard}},
  \bibinfo{author}{W.~{Bhimji}}, \bibinfo{author}{R.~{Al-Rfou}}, and
  \bibinfo{author}{Z.~{Luki{\'c}}}, \bibinfo{journal}{ArXiv e-prints}
  (\bibinfo{date}{Jun. 2017}), \eprint{1706.02390}.
\bibitem{Lanusse2018}
\bibinfo{author}{F.~{Lanusse}}, \bibinfo{author}{Q.~{Ma}},
  \bibinfo{author}{N.~{Li}}, \bibinfo{author}{T.~E. {Collett}},
  \bibinfo{author}{C.-L. {Li}}, \bibinfo{author}{S.~{Ravanbakhsh}},
  \bibinfo{author}{R.~{Mandelbaum}}, and \bibinfo{author}{B.~{P{\'o}czos}},
  \bibinfo{journal}{\mnras} \bibinfo{volume}{\textbf{473}},
  \bibinfo{pages}{3895} (\bibinfo{date}{Jan. 2018}), \eprint{1703.02642}.
\bibitem{Gunn1972}
\bibinfo{author}{J.~E. {Gunn}} and \bibinfo{author}{J.~R. {Gott}, III},
  \bibinfo{journal}{\apj} \bibinfo{volume}{\textbf{176}}, \bibinfo{pages}{1}
  (\bibinfo{date}{Aug. 1972}).
\bibitem{Press1974}
\bibinfo{author}{W.~H. {Press}} and \bibinfo{author}{P.~{Schechter}},
  \bibinfo{journal}{\apj} \bibinfo{volume}{\textbf{187}}, \bibinfo{pages}{425}
  (\bibinfo{date}{Feb. 1974}).
\bibitem{Bond1991}
\bibinfo{author}{J.~R. {Bond}}, \bibinfo{author}{S.~{Cole}},
  \bibinfo{author}{G.~{Efstathiou}}, and \bibinfo{author}{N.~{Kaiser}},
  \bibinfo{journal}{\apj} \bibinfo{volume}{\textbf{379}}, \bibinfo{pages}{440}
  (\bibinfo{date}{Oct. 1991}).
\bibitem{Sheth1999}
\bibinfo{author}{R.~K. Sheth} and \bibinfo{author}{G.~Tormen},
  \bibinfo{journal}{Monthly Notices of the Royal Astronomical Society}
  \bibinfo{volume}{\textbf{308}}(1), \bibinfo{pages}{119}
  (\bibinfo{date}{1999}).
\bibitem{Sheth2001}
\bibinfo{author}{R.~K. {Sheth}}, \bibinfo{author}{H.~J. {Mo}}, and
  \bibinfo{author}{G.~{Tormen}}, \bibinfo{journal}{\mnras}
  \bibinfo{volume}{\textbf{323}}, \bibinfo{pages}{1} (\bibinfo{date}{May
  2001}), \eprint{astro-ph/9907024}.
\bibitem{Han2018}
\bibinfo{author}{J.~{Han}}, \bibinfo{author}{Y.~{Li}},
  \bibinfo{author}{Y.~{Jing}}, \bibinfo{author}{T.~{Nishimichi}},
  \bibinfo{author}{W.~{Wang}}, and \bibinfo{author}{C.~{Jiang}},
  \bibinfo{journal}{ArXiv e-prints}  (\bibinfo{date}{Feb. 2018}),
  \eprint{1802.09177}.
\bibitem{Mao2018}
\bibinfo{author}{Y.-Y. {Mao}}, \bibinfo{author}{A.~R. {Zentner}}, and
  \bibinfo{author}{R.~H. {Wechsler}}, \bibinfo{journal}{\mnras}
  \bibinfo{volume}{\textbf{474}}, \bibinfo{pages}{5143} (\bibinfo{date}{Mar.
  2018}), \eprint{1705.03888}.
\bibitem{Lucie-Smith2018}
\bibinfo{author}{L.~{Lucie-Smith}}, \bibinfo{author}{H.~V. {Peiris}},
  \bibinfo{author}{A.~{Pontzen}}, and \bibinfo{author}{M.~{Lochner}},
  \bibinfo{journal}{ArXiv e-prints}  (\bibinfo{date}{Feb. 2018}),
  \eprint{1802.04271}.
\bibitem{Lowe2004}
\bibinfo{author}{D.~G. Lowe}, \bibinfo{journal}{International Journal of
  Computer Vision} \bibinfo{volume}{\textbf{60}}(2), \bibinfo{pages}{91}
  (\bibinfo{date}{Nov 2004}),
  \url{https://doi.org/10.1023/B:VISI.0000029664.99615.94}.
\bibitem{He2009}
\bibinfo{author}{H.~He} and \bibinfo{author}{E.~A. Garcia},
  \bibinfo{journal}{IEEE Trans. on Knowl. and Data Eng.}
  \bibinfo{volume}{\textbf{21}}(9), \bibinfo{pages}{1263} (\bibinfo{date}{Sep.
  2009}), \url{http://dx.doi.org/10.1109/TKDE.2008.239}.
\bibitem{Kingma2014}
\bibinfo{author}{D.~P. {Kingma}} and \bibinfo{author}{J.~{Ba}},
  \bibinfo{journal}{ArXiv e-prints}  (\bibinfo{date}{Dec. 2014}),
  \eprint{1412.6980}.
\bibitem{Chollet2015}
\bibinfo{author}{F.~Chollet} \emph{et~al.}, \bibinfo{title}{\emph{Keras}},
  \url{https://github.com/fchollet/keras} (\bibinfo{date}{2015}).
\bibitem{Abadi2016}
\bibinfo{author}{M.~{Abadi}}, \bibinfo{author}{A.~{Agarwal}},
  \bibinfo{author}{P.~{Barham}}, \bibinfo{author}{E.~{Brevdo}},
  \bibinfo{author}{Z.~{Chen}}, \bibinfo{author}{C.~{Citro}},
  \bibinfo{author}{G.~S. {Corrado}}, \bibinfo{author}{A.~{Davis}},
  \bibinfo{author}{J.~{Dean}}, \bibinfo{author}{M.~{Devin}}, \emph{et~al.},
  \bibinfo{journal}{ArXiv e-prints}  (\bibinfo{date}{Mar. 2016}),
  \eprint{1603.04467}.
\bibitem{Feng2016}
\bibinfo{author}{Y.~{Feng}}, \bibinfo{author}{M.-Y. {Chu}}, and
  \bibinfo{author}{U.~{Seljak}}, \bibinfo{journal}{ArXiv e-prints}
  (\bibinfo{date}{Mar. 2016}), \eprint{1603.00476}.
\bibitem{Hand2017}
\bibinfo{author}{N.~{Hand}}, \bibinfo{author}{Y.~{Feng}},
  \bibinfo{author}{F.~{Beutler}}, \bibinfo{author}{Y.~{Li}},
  \bibinfo{author}{C.~{Modi}}, \bibinfo{author}{U.~{Seljak}}, and
  \bibinfo{author}{Z.~{Slepian}}, \bibinfo{journal}{ArXiv e-prints}
  (\bibinfo{date}{Dec. 2017}), \eprint{1712.05834}.
\bibitem{Feng2016a}
\bibinfo{author}{Y.~{Feng}} and \bibinfo{author}{C.~{Modi}},
  \bibinfo{journal}{ArXiv e-prints}  (\bibinfo{date}{Jul. 2016}),
  \eprint{1607.03224}.
\bibitem{Rodriguez-Torres2016}
\bibinfo{author}{S.~A. {Rodr{\'{\i}}guez-Torres}}, \bibinfo{author}{C.-H.
  {Chuang}}, \bibinfo{author}{F.~{Prada}}, \bibinfo{author}{H.~{Guo}},
  \bibinfo{author}{A.~{Klypin}}, \bibinfo{author}{P.~{Behroozi}},
  \bibinfo{author}{C.~H. {Hahn}}, \bibinfo{author}{J.~{Comparat}},
  \bibinfo{author}{G.~{Yepes}}, \bibinfo{author}{A.~D. {Montero-Dorta}},
  \emph{et~al.}, \bibinfo{journal}{\mnras} \bibinfo{volume}{\textbf{460}},
  \bibinfo{pages}{1173} (\bibinfo{date}{Aug. 2016}), \eprint{1509.06404}.
\bibitem{Modi2017}
\bibinfo{author}{C.~{Modi}}, \bibinfo{author}{E.~{Castorina}}, and
  \bibinfo{author}{U.~{Seljak}}, \bibinfo{journal}{\mnras}
  \bibinfo{volume}{\textbf{472}}, \bibinfo{pages}{3959} (\bibinfo{date}{Dec.
  2017}), \eprint{1612.01621}.
\bibitem{Yoo2012}
\bibinfo{author}{J.~{Yoo}} and \bibinfo{author}{U.~{Seljak}},
  \bibinfo{journal}{\prd} \bibinfo{volume}{\textbf{86}}(8),
  \bibinfo{pages}{083504}, \bibinfo{eid}{083504} (\bibinfo{date}{Oct. 2012}),
  \eprint{1207.2471}.
\bibitem{Seo2010}
\bibinfo{author}{H.-J. Seo}, \bibinfo{author}{J.~Eckel}, \bibinfo{author}{D.~J.
  Eisenstein}, \bibinfo{author}{K.~Mehta}, \bibinfo{author}{M.~Metchnik},
  \bibinfo{author}{N.~Padmanabhan}, \bibinfo{author}{P.~Pinto},
  \bibinfo{author}{R.~Takahashi}, \bibinfo{author}{M.~White}, and
  \bibinfo{author}{X.~Xu}, \bibinfo{journal}{The Astrophysical Journal}
  \bibinfo{volume}{\textbf{720}}(2), \bibinfo{pages}{1650}
  (\bibinfo{date}{2010}), \url{http://stacks.iop.org/0004-637X/720/i=2/a=1650}.
\bibitem{Seo2016}
\bibinfo{author}{H.-J. {Seo}}, \bibinfo{author}{F.~{Beutler}},
  \bibinfo{author}{A.~J. {Ross}}, and \bibinfo{author}{S.~{Saito}},
  \bibinfo{journal}{\mnras} \bibinfo{volume}{\textbf{460}},
  \bibinfo{pages}{2453} (\bibinfo{date}{Aug. 2016}), \eprint{1511.00663}.
\bibitem{Seo2007}
\bibinfo{author}{H.-J. {Seo}} and \bibinfo{author}{D.~J. {Eisenstein}},
  \bibinfo{journal}{\apj} \bibinfo{volume}{\textbf{665}}, \bibinfo{pages}{14}
  (\bibinfo{date}{Aug. 2007}), \eprint{astro-ph/0701079}.
\bibitem{Tegmark1997}
\bibinfo{author}{M.~Tegmark}, \bibinfo{journal}{Phys. Rev. Lett.}
  \bibinfo{volume}{\textbf{79}}, \bibinfo{pages}{3806} (\bibinfo{date}{Nov
  1997}), \url{https://link.aps.org/doi/10.1103/PhysRevLett.79.3806}.
\bibitem{Noh2009}
\bibinfo{author}{Y.~{Noh}}, \bibinfo{author}{M.~{White}}, and
  \bibinfo{author}{N.~{Padmanabhan}}, \bibinfo{journal}{\prd}
  \bibinfo{volume}{\textbf{80}}(12), \bibinfo{pages}{123501},
  \bibinfo{eid}{123501} (\bibinfo{date}{Dec. 2009}), \eprint{0909.1802}.

\end{thebibliography}

\end{document}